\documentclass[twocolumn]{aastex63}

\usepackage{ulem}

\newcommand{\OII}{\textsc{{\rm O}\kern 0.1em{\sc ii}}}
\newcommand{\MgII}{\textsc{{\rm Mg}\kern 0.1em{\sc ii}}}
\newcommand{\FeII}{\textsc{{\rm Fe}\kern 0.1em{\sc ii}}}

\shorttitle{Circumgalactic \MgII{} in TNG and MEGAFLOW}
\shortauthors{DeFelippis et al.}

\graphicspath{{./}{figures/}}

\begin{document}

\title{A Comparison of Circumgalactic \MgII{} Absorption between the TNG50 Simulation and the MEGAFLOW Survey}

\correspondingauthor{Daniel DeFelippis}
\email{d.defelippis@columbia.edu}

\author[0000-0002-0112-7690]{Daniel DeFelippis}
\affiliation{Department of Astronomy, Columbia University, 550 West 120th Street, New York, NY 10027, USA}

\author[0000-0003-0068-9920]{Nicolas F. Bouch\'e}
\affiliation{Univ Lyon, Univ Lyon1, Ens de Lyon, CNRS, Centre de Recherche Astrophysique de Lyon UMR5574, F-69230 Saint-Genis-Laval, France}

\author[0000-0002-3185-1540]{Shy Genel}
\affiliation{Center for Computational Astrophysics, Flatiron Institute, 162 Fifth Avenue, New York, NY 10010, USA}
\affiliation{Columbia Astrophysics Laboratory, Columbia University, 550 West 120th Street, New York, NY 10027, USA}

\author[0000-0003-2630-9228]{Greg L. Bryan}
\affiliation{Department of Astronomy, Columbia University, 550 West 120th Street, New York, NY 10027, USA}
\affiliation{Center for Computational Astrophysics, Flatiron Institute, 162 Fifth Avenue, New York, NY 10010, USA}

\author[0000-0001-8421-5890]{Dylan Nelson}
\affiliation{Universit\"at Heidelberg, Zentrum f\"ur Astronomie, Institut f\"ur theoretische Astrophysik, Albert-Ueberle-Str. 2, D-69120 Heidelberg, Germany}

\author[0000-0003-3816-7028]{Federico Marinacci}
\affiliation{ Department of Physics and Astronomy ``Augusto Righi,'' University of Bologna, Via Gobetti 93/2, I-40129 Bologna, Italy}

\author[0000-0001-6950-1629]{Lars Hernquist}
\affiliation{Harvard-Smithsonian Center for Astrophysics, 60 Garden Street, Cambridge, MA 02138, USA}

\begin{abstract}

The circumgalactic medium (CGM) contains information on gas flows around galaxies, such as accretion and supernova-driven winds, which are difficult to constrain from observations alone. Here, we use the high-resolution TNG50 cosmological magnetohydrodynamical simulation to study the properties and kinematics of the CGM around star-forming galaxies in $10^{11.5}-10^{12}\;M_{\odot}$ halos at $z\simeq1$ using mock \MgII{} absorption lines, which we generate by postprocessing halos to account for photoionization in the presence of a UV background. We find that the \MgII{} gas is a very good tracer of the cold CGM, which is accreting inward at inflow velocities of up to $50\;\rm{km\;s^{-1}}$. For sight lines aligned with the galaxy's major axis, we find that \MgII{} absorption lines are kinematically shifted due to the cold CGM's significant corotation at speeds up to 50\% of the virial velocity for impact parameters up to 60 kpc. We compare mock \MgII{} spectra to observations from the MusE GAs FLow and Wind (MEGAFLOW) survey of strong \MgII{} absorbers ($\rm{EW}^{2796\AA}_{0}>0.5$ \AA). After matching the equivalent-width (EW) selection, we find that the mock \MgII{} spectra reflect the diversity of observed kinematics and EWs from MEGAFLOW, even though the sight lines probe a very small fraction of the CGM. \MgII{} absorption in higher-mass halos is stronger and broader than in lower-mass halos but has qualitatively similar kinematics. The median-specific angular momentum of the \MgII{} CGM gas in TNG50 is very similar to that of the entire CGM and only differs from non-CGM components of the halo by normalization factors of $\lesssim1\;\rm{dex}$.

\end{abstract}


\section{Introduction} \label{sec:intro}

The accretion of gas onto disk galaxies is a fundamental part of galaxy formation and evolution, as gas within disks is continually used to form stars and must therefore be regularly replenished \citep[e.g.,][]{Putman17}. All such gas, whether pristine gas from cosmological inflows or recycled gas in the process of reaccreting, must pass through the local environment surrounding galaxies, often called the circumgalactic medium (CGM). The CGM might contain a substantial amount of angular momentum as shown by many studies of galaxy simulations \citep[e.g.,][]{Stewart11,Danovich15,DeFelippis20}. As the gas accretes onto the galaxy, the angular momentum will flow inward too, meaning the CGM is a source not just of the mass of the disk, but its angular momentum as well. 

Not all gas surrounding galaxies is inflowing though: the CGM also contains outflowing gas ejected from the galaxy by feedback from supernovae and active galactic nuclei (AGN), which is capable of affecting the way in which CGM gas eventually joins the galaxy \citep{DeFelippis17}. All of these physical processes occur concurrently and result in a multiphase environment shown in observations to have complex kinematics \citep[see][and references therein]{Tumlinson17}.

A large number of recent observations of the CGM have been accomplished through absorption line studies of background quasars through dedicated surveys \citep[e.g.,][]{Liang14,Borthakur15,Kacprzak15}. For instance, some surveys are constructed by cross-correlating quasar absorption lines with spectroscopic redshift surveys such as the Keck Baryonic Structure Survey \citep[KBSS;][]{Rakic12,Rudie12,Turner14} or with photometric surveys like the Sloan Digital Sky Survey \citep[SDSS;][]{Huang16,Lan18,Lan20}. Other CGM surveys attempt to either match individual absorption lines to known galaxies (i.e.,~are ``galaxy selected''), like the COS-Halos \citep[e.g.,][]{Tumlinson11,Werk13,Borthakur16,Burchett19}, COS-LRG \citep{Chen18,Zahedy19}, and the low-redshift Keck surveys conducted at Keck Observatory \citep{Ho17,MartinC19}, or match galaxies near known absorbers (i.e.,~``absorber selected'') such as the MusE GAs FLOw and Wind survey \citep[MEGAFLOW;][]{Schroetter16,Schroetter19,Schroetter21,Wendt21,Zabl19,Zabl20,Zabl21}. In these surveys, there is generally only one quasar sight line per galaxy, but in certain rare cases it is possible to find multiple sight lines associated with a single galaxy through multiple quasars \citep{Bowen16}, a single multiply lensed quasar \citep{Chen14,Zahedy16,Kulkarni19}, an extended lensed quasar \citep{Lopez18}, or even an extended background galaxy \citep{Diamond-Stanic16}.

The \MgII{} ion has been a focus of many recent surveys including the \MgII{} Absorber-Galaxy Catalog \citep[MAGIICAT;][]{Chen08,Chen10a,Nielsen13a,Nielsen13b,Nielsen15}, the Magellan MagE \MgII{} (M3) Halo Project \citep{Chen10a,Chen10b,Huang21}, the MUSE Analysis of Gas around Galaxies Survey \citep[MAGG;][]{Dutta20}, the PRIsm MUlti-object Survey \citep[PRIMUS;][]{Coil11,Rubin18}, and the aforementioned MEGAFLOW survey, as well as individual absorbers \citep[e.g.,][]{Lopez20}. These studies belong to a long history of \MgII{} $\lambda2796$ absorption line surveys \citep[e.g.,][]{Bergeron91,Bergeron92,Steidel92}, which unveiled the first galaxy--absorber pairs at intermediate redshifts. Though not the focus of this paper, \MgII{} has also been seen in emission in extended structures around the galaxy and in the CGM \citep[e.g.,][]{Rubin11,RickardsVaught19,Rupke19,Burchett21,Zabl21}.

Along with this wealth of \MgII{} observations, researchers in recent years have found \MgII{} kinematics to be correlated over large spatial scales. In particular, both \cite{Bordoloi11} and \cite{Bouche12} found a strong dependence of \MgII{} absorption on azimuthal angle: specifically, more absorption near $\phi=0^{\circ}$ and $90^{\circ}$ and a lack of absorption near $45^{\circ}$. This type of absorption distribution is generally interpreted as bipolar outflows along the minor axis and inflows along the major axis. In this context, both galaxy-selected \citep[e.g.,][]{Ho17,MartinC19} and absorption-selected \MgII{} studies \citep[e.g.,][]{Kacprzak12,Bouche13,Bouche16,Zabl19} have given support to the interpretation of accretion of gas from the CGM onto the galaxy. These \MgII{} studies show that when sight lines are located near the major axis of the galaxy there are clear signatures of corotating cold gas with respect to the galaxy kinematics.

However, despite such extensive observational data, developing a general understanding of cold gas in the CGM from the \MgII{} line alone remains difficult due to the limited spatial information provided by the observational technique (though IFU mapping of lensed arcs in \citealp[e.g.,][]{Lopez20}, \citealp{Mortensen21}, and \citealp{Tejos21} can improve this in the future), as well as the fact that \MgII{} gas may not be representative of the entire cold phase of the CGM. To study more physically fundamental properties of the CGM, it is therefore necessary to turn to galaxy simulations. 

In cosmological simulations (see \citealp{Vogelsberger20} for a review), the CGM has been notoriously difficult to model accurately due to the need to resolve very small structures \citep[e.g.,][]{Hummels19,Peeples19,Suresh19,Corlies20}. Nonetheless, the CGM has been shown to preferentially align with and rotate in the same direction of the galaxy, especially near the galaxy's major axis \citep{Stewart13,Stewart17,Ho19,DeFelippis20}, which is  qualitatively consistent with observations in the same spatial region of the CGM \citep[e.g.,][]{Zabl19}. However, this general qualitative agreement between simulations and observations is difficult to put on firm ground quantitatively due to the inherent differences between observations and simulations.

In this paper, we analyze a set of halos from the TNG50 simulation \citep{Nelson19,Pillepich19} using the \textsc{Trident} tool \citep{Hummels17} to model the ionization state of the CGM and then perform a quantitative comparison of the kinematics of the cool ($T \lesssim 3\times10^4 \; \rm{K}$) CGM traced by \MgII{} absorption to major-axis sight lines from the MEGAFLOW survey \citep{Zabl19} while attempting to match the observational selection criteria as described in Section \ref{sec:methods}. We note that our comparison to MEGAFLOW galaxies with stellar masses $M_{*} \sim 10^{10} \; M_{\odot}$ is complementary to that of both \cite{Nelson20}, who study the origins of cold CGM gas of very massive galaxies ($M_{*} > 10^{11} \; M_{\odot}$), and \cite{Nelson21}, who study properties of extended \MgII{} emission in the CGM. 

The paper is organized as follows. In Section \ref{sec:methods}, we describe the TNG50 simulation and MEGAFLOW sample used in the comparison, and we outline the analysis pipeline used to generate mock observations. In Section \ref{sec:results}, we describe our main results, first by comparing the simulated and real observations, then by analyzing the features of the simulation that give rise to the properties of the mock observations. In Section \ref{sec:discussion}, we discuss the implications of our results for the role of the CGM in galaxy formation, and we summarize our findings in Section \ref{sec:summary}.

\section{Methods} \label{sec:methods}

\subsection{Simulations}

We utilize the TNG50 simulation \citep{Nelson19,Pillepich19}, the highest-resolution version of the IllustrisTNG simulation suite \citep{Marinacci18,Naiman18,Nelson18,Pillepich18,Springel18}, which is itself based on the original Illustris simulation \citep{Vogelsberger14a,Vogelsberger14b}. TNG50 evolves a periodic $\approx (52 \; \rm{Mpc})^3$ box from cosmological initial conditions to $z=0$ with the moving-mesh code \textsc{Arepo} \citep{Springel10,Weinberger20}. It has a baryonic mass resolution of $\sim8.5\times10^4 \; M_{\odot}$ per cell, which is a factor of $\approx 16$ better than the resolution of TNG100. We discuss the effect of simulation resolution on our results later in Section \ref{sec:results}.

\subsection{Observational Data}
\label{sec:methods2}

The MEGAFLOW survey (N. Bouch\'e et al., in preparation) consists of a sample of 79 \MgII{} $\mathrm{\lambda\lambda}2796,2803$ absorbers in 22 quasar lines of sight observed with the Multi-Unit Spectroscopic Explorer \citep[MUSE;][]{Bacon06}. The quasars were selected to have at least three \MgII{} absorbers from the \cite{Zhu13} SDSS catalog in the redshift range $0.4 < z < 1.4$ such that the [\OII] $\lambda\lambda 3727,3729$ galaxy emission lines fell within the MUSE wavelength range ($4800-9300 \; \rm{\AA}$). A threshold on the rest-frame equivalent width of $\sim0.5-0.8 \; \rm{\AA}$ was also imposed on each absorber. 

For this paper, we focus on a preliminary subset of the MEGAFLOW sample of \MgII{} absorber--galaxy pairs whose quasar location is positioned within 35$^\circ$ of the major axis of the host galaxy \citep[][]{Zabl19}. This subset consists of nine absorber--galaxy pairs with redshifts $0.5 < z < 1.4$ and impact parameters ($b$) ranging from 13 to 65 kpc with a mean of $\approx34$ kpc. \citet{Zabl19} found that the \MgII{} gas in these absorbers show a strong preference for corotation with their corresponding host galaxies.

The galaxies in \citet{Zabl19} are both fairly isolated by having at most one companion within $100 \; \rm{kpc}$, and star forming with [\OII] fluxes $f_{\OII}>4\times 10^{-17} \; \rm{erg \; s^{-1} \; cm^{-2}}$, i.e., star-formation rates $\gtrsim 1 \; M_{\odot} \; \rm{yr^{-1}}$. The galaxies have stellar masses $M_{*}$ ranging from $10^{9.3}-10^{10.5} \; M_{\odot}$ and halo masses $M_{\rm{vir}}$ ranging from $\approx 10^{11.4}-10^{12.2} \; M_{\odot}$, where $M_{\rm{vir}}$ is defined from the stellar mass--halo mass relation from \cite{Behroozi10}. As \cite{Zabl19} show, these halo masses generally match the \cite{Bryan98} definition of $M_{\rm{vir}}$. 

\subsection{Sample selection and Forward Modeling}
\label{sec:methods3}

\begin{figure}
\fig{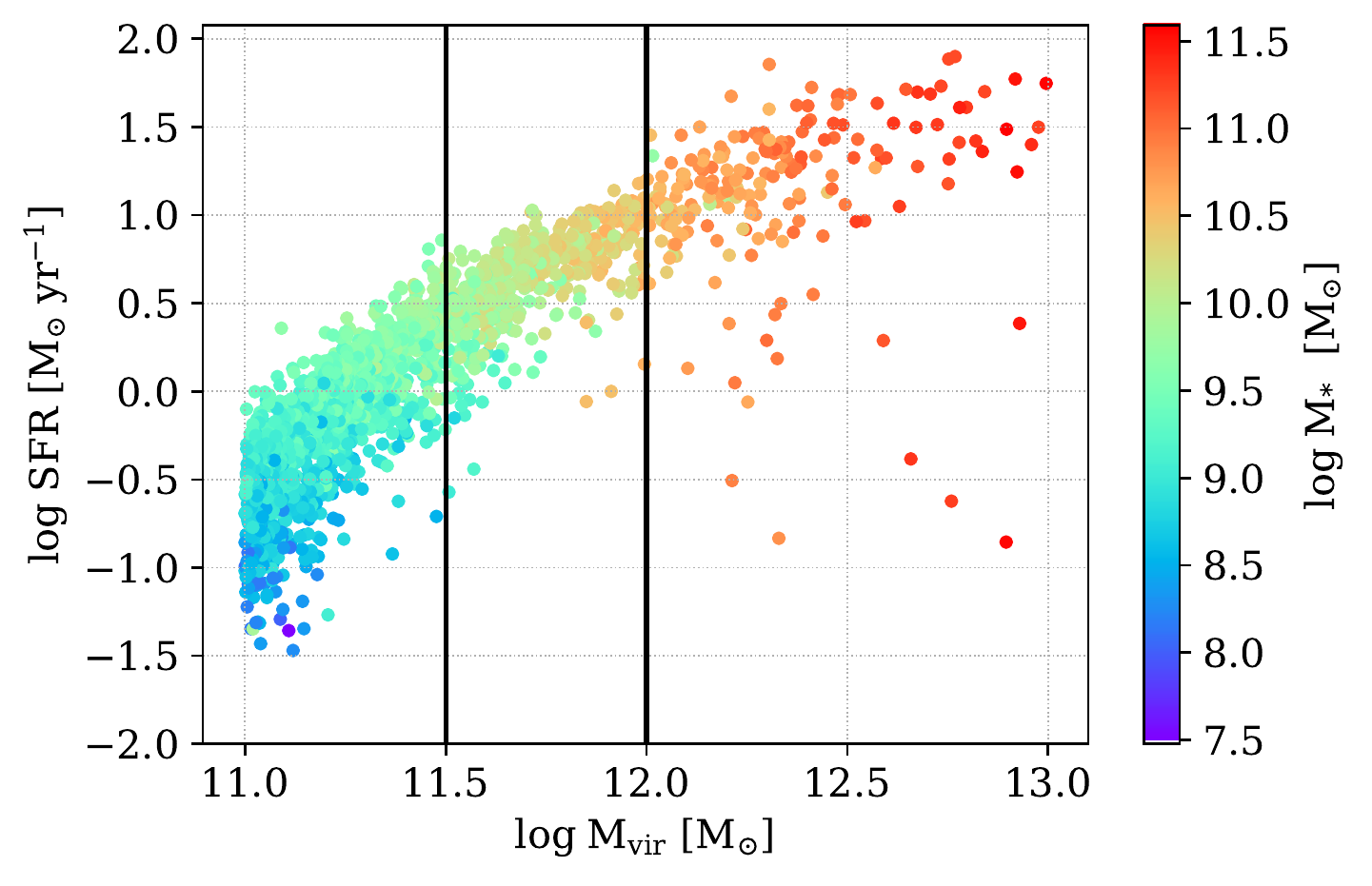}{0.48\textwidth}{}
\vspace{-25pt}
\caption{Star formation rate of the central galaxy vs. halo mass for all TNG50 halos between $10^{11} \; M_{\odot}$ and $10^{13} \; M_{\odot}$ at $z=1$. Each point is colored by the stellar mass of the halo's central galaxy. Two thick vertical lines demarcate the halo mass range of the fiducial sample.
}
\label{f:SFR_vs_Mvir}
\end{figure}

Figure \ref{f:SFR_vs_Mvir} shows the central galaxies' instantaneous star formation rates (SFR) and stellar masses of all TNG50 halos in and around the mass range of interest. Since we aim to compare the \MgII{} absorption properties of mock line-of-sight (LOS) observations through TNG50 halos to those of major-axis sight lines of the MEGAFLOW survey, we first select a sample of simulated halos at $z=1$ in the mass range $10^{11.5} \; M_{\odot} < M_{\rm{halo}} < 10^{12} \; M_{\odot}$ using the \cite{Bryan98} definition for $M_{\rm{halo}}$, which results in a sample of 495 halos. In the remainder of this paper, we will refer to this subsample as the ``fiducial'' sample. The chosen redshift is typical for the \citet{Zabl19} sample, and the halo mass range covers the typical inferred virial masses of their halos. Nearly all of the halos in our fiducial sample host central galaxies with SFR $\gtrsim1 \; \rm{M_{\odot} \; yr^{-1}}$ and stellar masses of $\sim 10^{10} \; \rm{M_{\odot}}$, which is consistent with the MEGAFLOW subsample as described in Section \ref{sec:methods2}.

For each halo, we adjust all velocities to be in the center-of-mass frame of the stars in the central galaxy, and we rotate it so that the stellar specific angular momentum of the central galaxy points in the $+z$-direction (the $x$- and $y$-directions are both arbitrary). With this geometry we then define a sight line in the $x-z$ plane by the impact parameter $b$, the azimuthal angle $\alpha$, and the inclination angle $i$, where $b$ is the projected distance from the center of the galaxy in the $y-z$ plane (i.e.,~``sky'' plane), $\alpha$ is the angle above the rotational plane of the galaxy, and $i$ is the angle of the sight line with respect to the sky plane. In this setup, edge-on and face-on views have $i=90^{\circ}$ and $i=0^{\circ}$, respectively (see Figure 1 of \citealp{Zabl19} for a sketch of the geometry described here). In order to mimic the observations of \citet{Zabl19}, we select sight lines through each halo at values of $b$ ranging from $15 \; \rm{kpc}$ to $60 \; \rm{kpc}$, $\alpha = 5^{\circ}$ and $25^{\circ}$, and at $i = 60^{\circ}$, representing the average inclination angle of a random sight line.

In order to generate observations of our TNG50 sample, we use the \textsc{Trident} package \citep{Hummels17}, which calculates ionization parameters for outputs of galaxy simulations using properties of the simulated gas cells and \textsc{Cloudy} \citep{Ferland13} ionization tables. These tables take as input the gas temperature, density, metallicity, and cosmological redshift of each gas cell and provide ionization fractions and number densities of desired ions. We make use of the current development version of \textsc{Trident}\footnote{\url{http://trident-project.org}} (v1.3), which itself depends on the current development version of \textsc{yt}\footnote{\url{https://yt-project.org}} (v4.0). In this paper, we use a set of ion tables created assuming collisional ionization equilibrium, photoionization from a \cite{Faucher-Giguere09} UV background, and self-shielding of neutral hydrogen (for details see \citealp{Emerick19} and \citealp{Li21}), as this was the background radiation model used to evolve the TNG50 simulation. We also use the elemental abundance of magnesium in each gas cell tracked by the simulation rather than assuming a constant solar abundance pattern throughout the halo to achieve greater self-consistency with TNG50. We note, however, that our results are not particularly sensitive to either of these choices. 

\begin{figure}
\fig{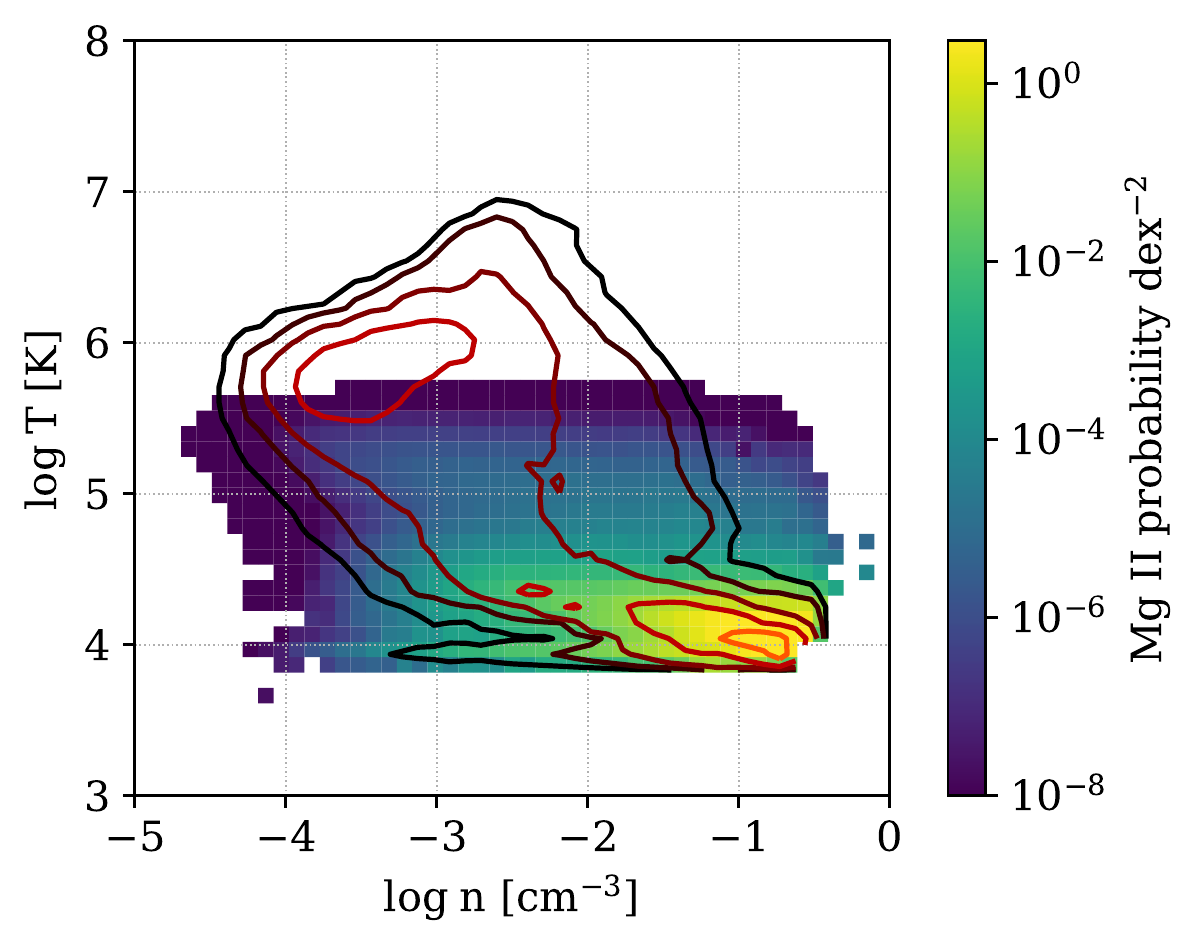}{0.48\textwidth}{}
\vspace{-25pt}
\caption{Temperature--number density phase diagram of a single TNG50 halo at $z=1$, colored by the \MgII{} mass probability density per dex$^2$. Contours show the distribution of all gas mass in the halo. }
\label{f:MgII_phase}
\end{figure}

Since our focus is on the \MgII{} $\lambda 2796$ line,
we show in Figure \ref{f:MgII_phase} a temperature--density phase diagram of the gas in one of the TNG50 halos from our sample, colored by the \MgII{} mass probability density. From this plot, it is clear that \MgII{} is mostly formed from the coldest ($\lesssim 10^{4.5} \; \rm{K}$) and densest ($\gtrsim 0.01 \; \rm{cm^{-3}}$) gas in the halo, though some \MgII{} mass exists at a larger range of temperatures and densities. However, contours showing the total gas mass demonstrate that despite this large range in temperature and density, essentially none of the diffuse ``hot'' phase, comparable in mass to the cold phase, contributes to \MgII{} absorption. We also note here that for this analysis we are excluding star-forming gas as its temperature and density are defined using an effective equation of state \citep{Springel03} and are therefore not analogous to the physical properties of non-star-forming gas. Properly modeling the physical properties of the star-forming gas (see \citealp{Padilla21} for an example of this technique) introduces a level of complexity not necessary for this analysis: we find that our results are not affected by the exclusion of this gas since our sight lines through the CGM rarely intersect any star-forming gas cells as most of them are within the galactic disk. 

\section{Results} \label{sec:results}

\begin{figure*}
\centering
\gridline{
\fig{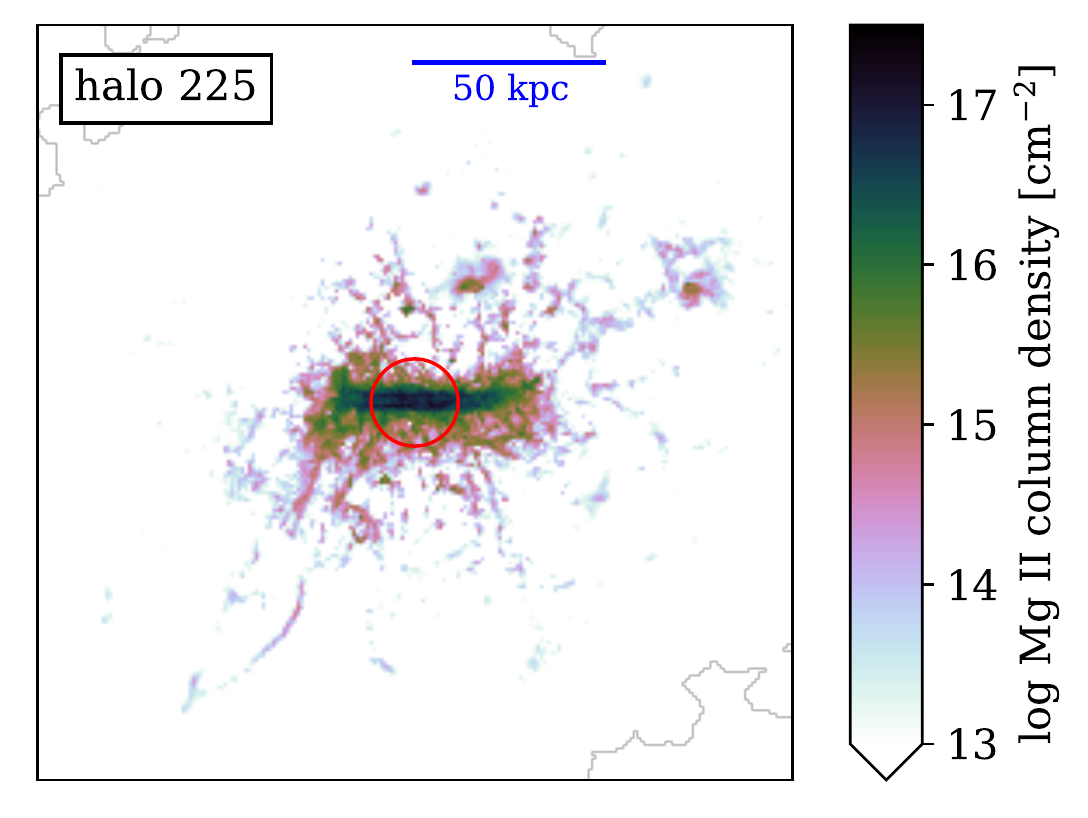}{0.6\textwidth}{}
}
\vspace{-25pt}
\gridline{
\fig{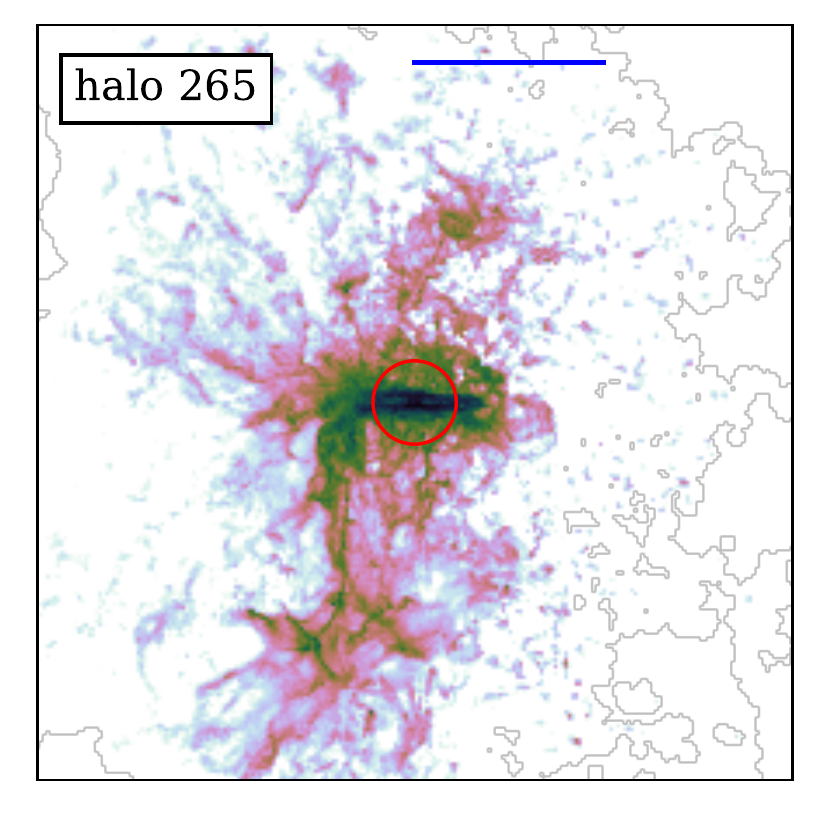}{0.33\textwidth}{}
\fig{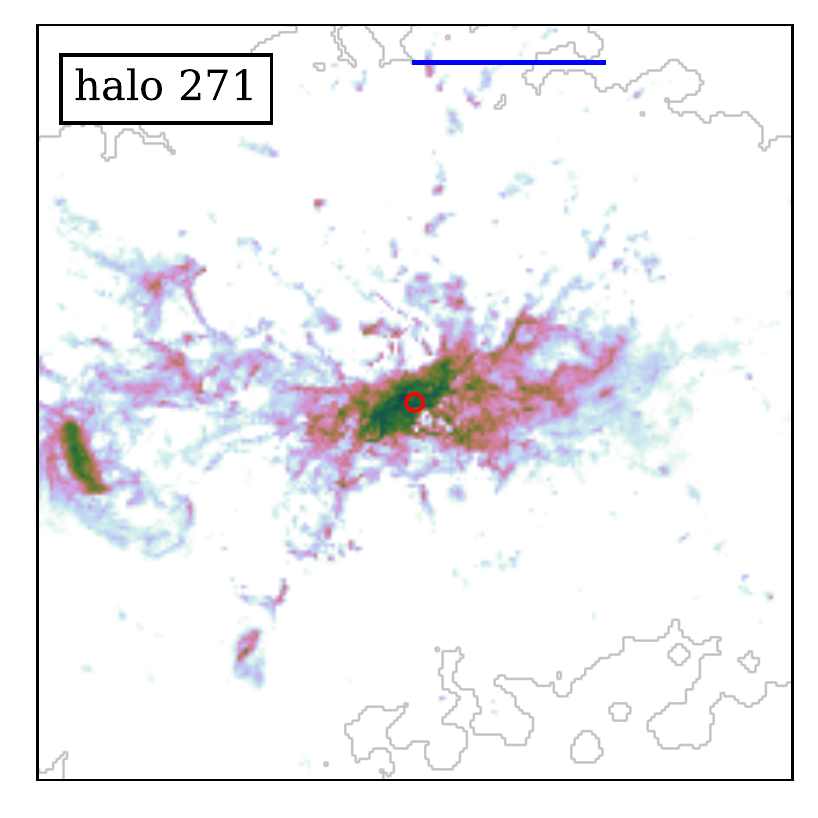}{0.33\textwidth}{}
\fig{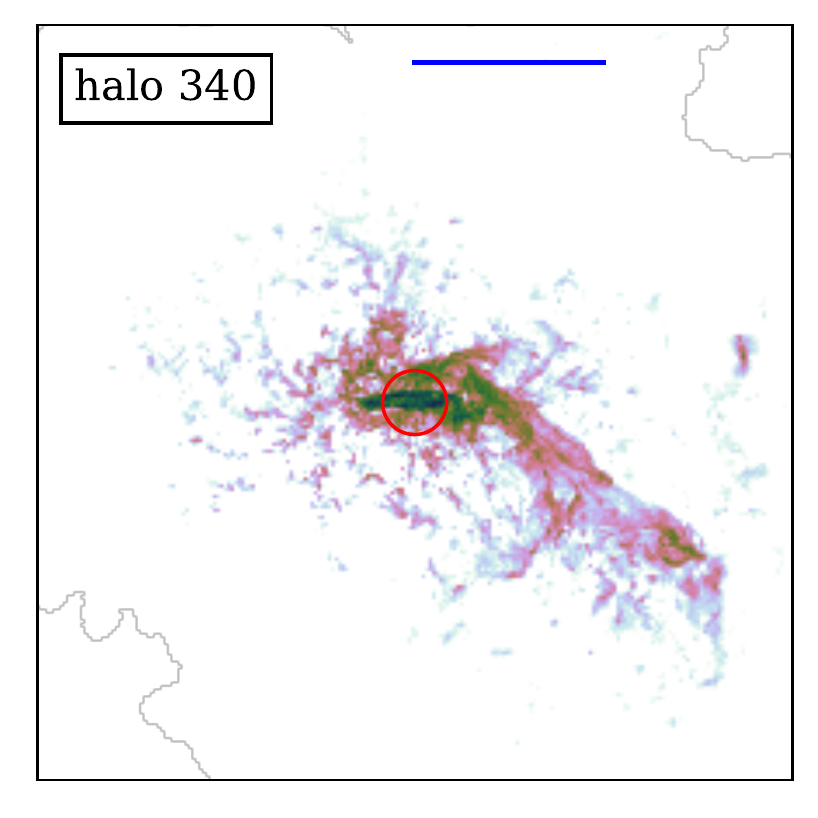}{0.33\textwidth}{}
}
\vspace{-25pt}
\caption{\MgII{} column density maps of four TNG50 halos from the fiducial halo mass bin of $10^{11.5} \; M_{\odot} < M_{\rm{halo}} < 10^{12} \; M_{\odot}$ at $z=1$, aligned so the angular momentum vector of the stars in the central galaxy points along the vertical axis (i.e., edge-on). The lower limit of the color bar is chosen to approximate observational detection limits. The red circle in each panel is centered on the galaxy and has a radius of twice the galaxy's stellar half-mass--radius, and the blue scale-bar shows a distance of $50 \; \rm{kpc}$ on the maps. The complexity and diversity of \MgII{} structure in the CGM of similar-mass halos are evident even in this small sample.
}
\label{f:MgII_maps}
\end{figure*}

We first present in Section \ref{sec:results1} the results of directly comparing the \MgII{} properties of TNG50 and MEGAFLOW using the analysis described in Section \ref{sec:methods}. Then, we further analyze the 3D kinematic properties of the \MgII{}-bearing gas from TNG50 in Section \ref{sec:results2} and consider evolution of \MgII{} absorption properties with halo mass and simulation resolution in Section \ref{sec:results3}. 

\subsection{Comparing TNG50 to MEGAFLOW}
\label{sec:results1}

In Figure \ref{f:MgII_maps}, we show \MgII{} column density maps of a selection of TNG50 halos drawn from our fiducial sample at $z=1$. The halos are aligned so that the angular momentum vector of the stars in the central galaxy points along the vertical axis; thus, the view is edge-on. The strongest \MgII{} columns are found within and very close to the galaxy, demarcated by a red circle with a radius of twice the galaxy's stellar half-mass--radius \citep[the same definition used in][]{DeFelippis20}. Beyond the galaxy, \MgII{} gas consistently appears to both surround the galactic disk and be very clumpy, but the amount and morphology of such gas varies greatly. In particular, there is significant variation with azimuthal angle: the highest \MgII{} columns generally appear in the plane of rotation, but strong columns can occur above and below the disk as well, such as in halo 265 (the bottom left panel of Figure~\ref{f:MgII_maps}). \citet{Peroux20} found the CGM gas metallicity to vary with azimuthal angle, but interestingly, they found gas near the major axis to have lower than average metallicity in the halo, indicating that large \MgII{} columns do not necessarily correspond to metal-enriched gas. High \MgII{} columns are much less common in the outer halo ($r \gtrsim 50 \; \rm{kpc}$), but the presence of satellite galaxies can populate that region with \MgII{} gas, shown most clearly in halo 340 (the bottom right panel of Figure \ref{f:MgII_maps}). 

Within our fiducial sample, it is evident that the distribution of \MgII{} varies drastically, presumably due to different halo formation histories. Sight lines through different halos will therefore likely produce different absorption profiles even for sight lines with identical geometries. This highlights the necessity of calculating population averages of \MgII{} properties from TNG50 to compare to MEGAFLOW.

\begin{figure}
\fig{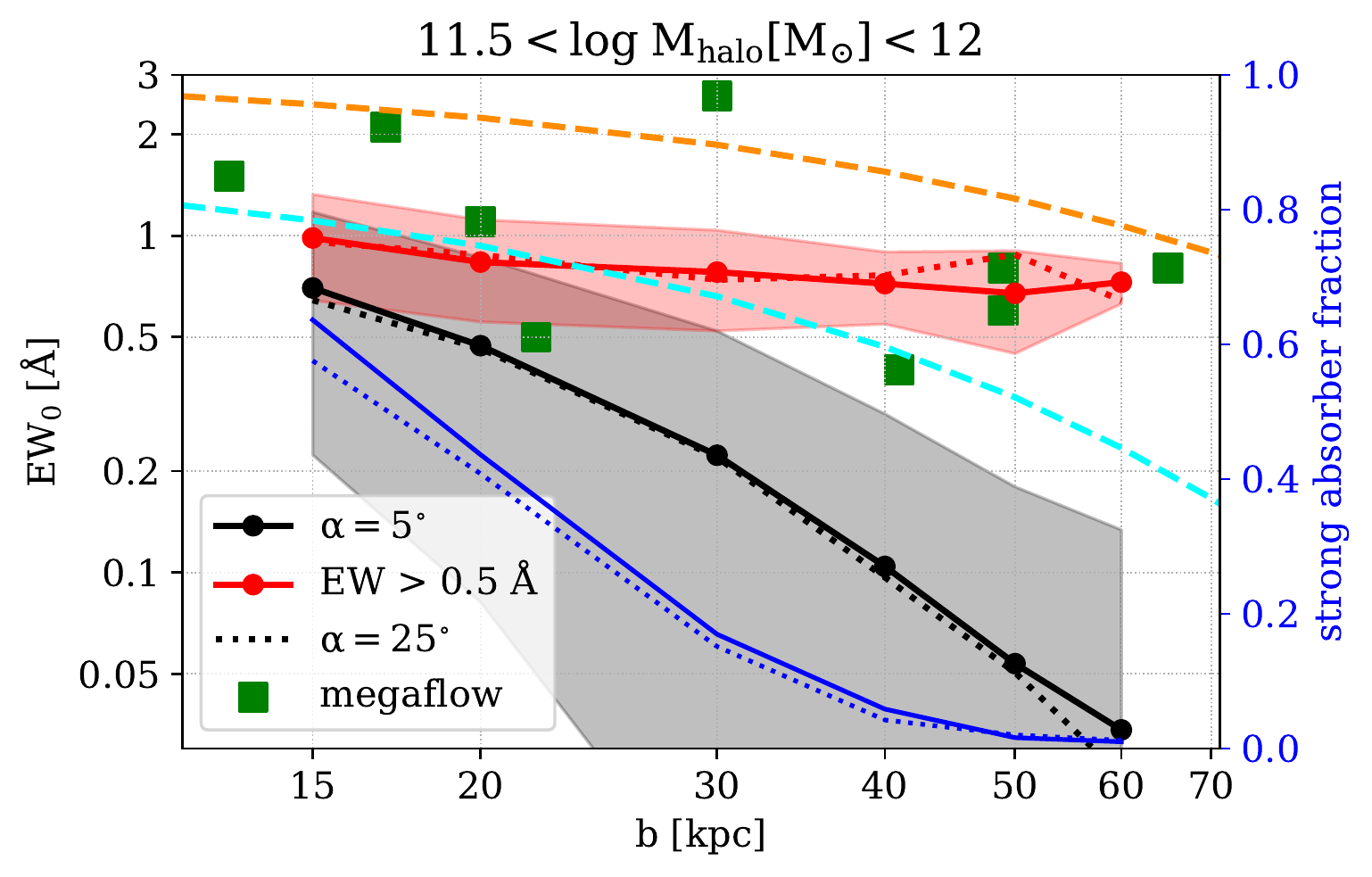}{0.48\textwidth}{}
\vspace{-25pt}
\caption{Mean \MgII{} equivalent widths of halos in our fiducial sample vs. the impact parameter of sight lines through those halos. Black and red lines and corresponding shaded regions show the mean and $\pm1\sigma$ scatter of all halos and the subset of strong absorbers (EW$_0 > 0.5 \; \rm{\AA}$), respectively. Sight lines at a constant azimuth angle of $\alpha = 5^{\circ}$ and $25^{\circ}$ are shown with solid and dotted lines, respectively. Observations of individual accretion systems from \citet{Zabl19} are shown as green squares. The fraction of strong absorbers as a function of impact parameter (blue) is shown with the right vertical axis. The cyan and orange dashed lines are log-linear fits of $z\sim1$ \MgII{} absorbers from \cite{Nielsen13b} and \cite{Lundgren21}, respectively.
}
\label{f:EW_vs_b}
\end{figure}

We begin such a comparison with Figure \ref{f:EW_vs_b}, which shows the average strength of \MgII{} absorption, represented as the rest-frame equivalent width (EW$_0$) as a function of impact parameter ($b$) for our fiducial sample. In this plot, we make an important distinction between the entire fiducial sample, shown in black, and the subset of ``strong absorbers'' in red. We define strong absorbers as sight lines through a halo that produce an absorption spectrum with EW$_0 > 0.5 \; \rm{\AA}$ \citep[the same as in][]{Zabl19}. It is this ``absorber-selected'' subset of the fiducial sample that is most directly comparable to MEGAFLOW. For easier comparison to Figure \ref{f:MgII_maps}, we find that sight lines with EW$_0 = 0.5 \; \rm{\AA}$ have \MgII{} column densities ranging from $\approx10^{13.5}-10^{14.5} \; \rm{cm^{-2}}$, i.e., just above the lower limit of the color bar.

At all impact parameters, the average rest-frame EW of the ``all absorbers'' sample from TNG50 (black) is smaller than those of MEGAFLOW, as expected given the selection function. The difference ranges from a factor of only $\approx 3$ at $b \leq 20 \; \rm{kpc}$ to a factor of $\approx 30$ at $60 \; \rm{kpc}$. If, instead, we compare the average EW$_0$ of the strong absorber subset (EW$_0 > 0.5 \; \rm{\AA}$) from TNG50, which is the appropriate comparison to make, we find the mean shown in red. This is much more similar to the values from MEGAFLOW, especially for $b \geq 40 \; \rm{kpc}$, but it is still as much as a factor of $\approx 2$ lower than the observed values at $b \leq 20 \; \rm{kpc}$. However, the limited size and large scatter of the MEGAFLOW points from \citet{Zabl19} make it difficult to assess the precise level of disagreement with TNG50. Sight lines at $\alpha = 5^{\circ}$ (solid) and $\alpha = 25^{\circ}$ (dotted) produce essentially identical equivalent widths over both the entire fiducial sample and the subset of strong absorbers. With the two additional dashed lines in Figure \ref{f:EW_vs_b} we provide a point of comparison to larger samples of moderate-redshift \MgII{} absorbers from \cite{Nielsen13b} and \cite{Lundgren21}. Though both of these samples have a slightly smaller equivalent-width threshold than \cite{Zabl19} ($\approx 0.2-0.3 \; \rm{\AA}$) and no selection based on the geometry of the sight line, they still bracket both the \cite{Zabl19} absorbers and the strong absorbers from TNG50, indicating that these simulated \MgII{} EWs are also consistent with observed \MgII{} EWs in general, given the large scatter.

The blue lines in Figure \ref{f:EW_vs_b} show the fraction of all sight lines that host strong absorbers as a function of impact parameter. At sight lines very close to the galaxy ($b = 15 \; \rm{kpc}$), strong absorbers are common and in fact represent a majority of all halos. However, by $b = 20 \; \rm{kpc}$ the strong absorber fraction drops below 50\%, and at the largest impact parameters shown, the fraction is only $\approx 1\%$. Strong absorbers are slightly more common at $\alpha = 5^{\circ}$ compared to $\alpha = 25^{\circ}$, which can be understood by noting that the sight lines with smaller $\alpha$ pass through the disk midplane closer to the galaxy's center, where gas is generally denser. However, this difference in strong absorber fraction does not affect the measured equivalent widths, indicating that the TNG50 halos' agreement with MEGAFLOW for sight lines near the galaxies' major axes is not subject to the precise geometries of the sight lines. 

\begin{figure}
\fig{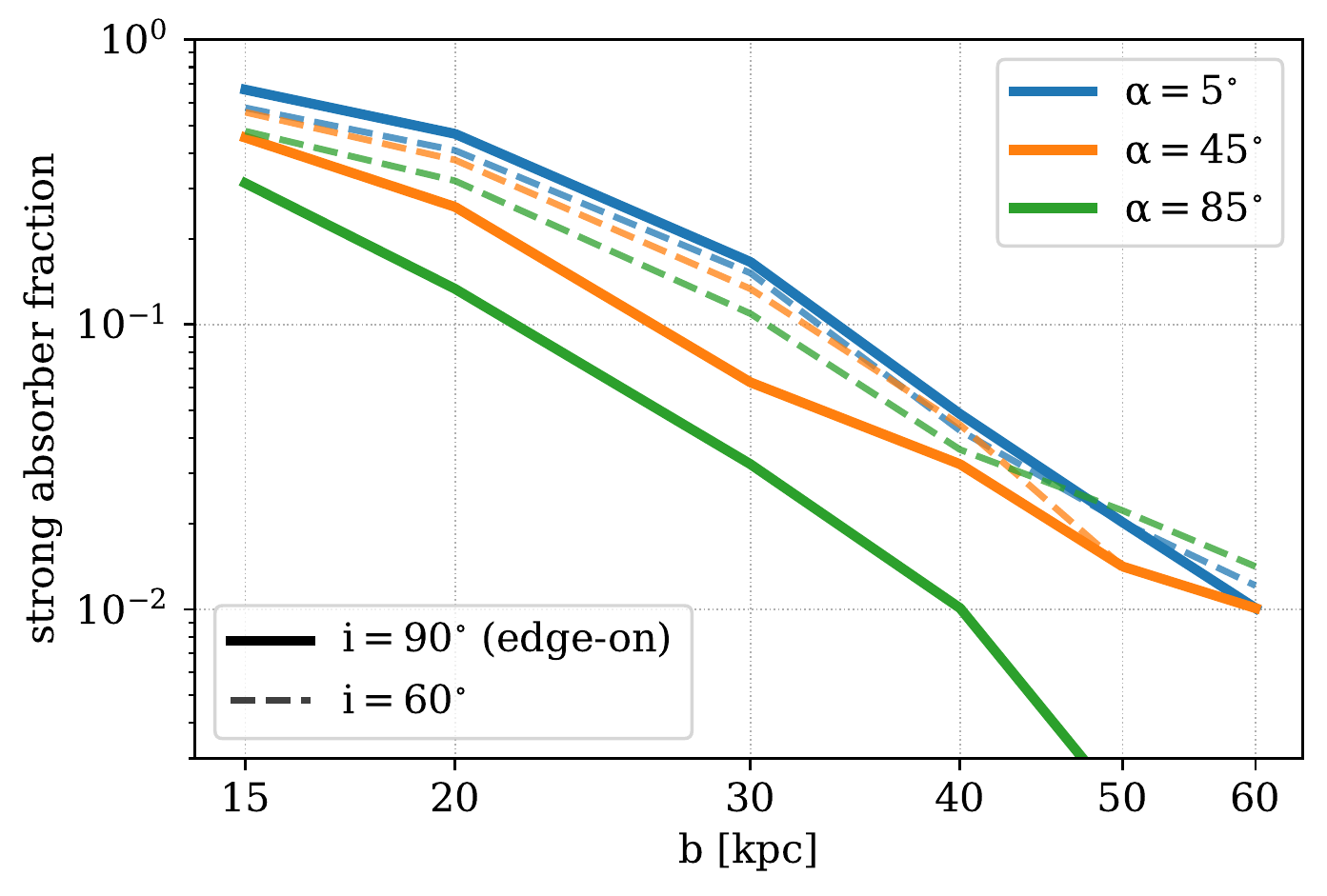}{0.48\textwidth}{}
\vspace{-25pt}
\caption{\MgII{} covering fraction of our fiducial sample as a function of impact parameter for sight lines  with $\alpha=5^{\circ}$ (blue), $\alpha=45^{\circ}$ (orange), and $\alpha=85^{\circ}$ (green). Solid and dashed lines show sight lines that are edge-on and inclined at $i=60^{\circ}$, respectively.
}
\label{f:covering}
\end{figure}

In Figure \ref{f:covering}, we examine how \MgII{} EWs vary throughout the entire halo in TNG50, not just near the major axis, and we find a clear trend: at all impact parameters we study, the mean EW of a perfectly edge-on sight line decreases as the azimuth angle of that sight line $\alpha$ increases. Sight lines near the minor axis (green) have EWs at least $0.35 \; \rm{dex}$ smaller than sight lines near the major axis (blue), and sight lines between both axes (orange) have EWs between the values at both axes. This represents a disagreement between TNG50 and \MgII{} observations, which are generally observed to have a bimodal distribution of $\alpha$ near $0^{\circ}$ and $90^{\circ}$ \citep{Bordoloi11,Bouche12,Kacprzak12,MartinC19,Zabl19,Lundgren21}. The distribution of $\alpha$ in TNG50 is clearly peaked at small $\alpha$, implying that TNG50 is not producing the same kind of \MgII{} that is inferred to be outflowing in observations. It is also clear that this azimuthal angle dependence is very sensitive to the inclination angle of the sight line because it nearly disappears when the sight lines are inclined at an angle of $60^{\circ}$ with respect to the axis of rotation (dotted lines in Figure \ref{f:covering}), as would be typical for observations. This sensitivity indicates that most \MgII{} absorption in TNG50 comes from a gas in the vicinity of the disk midplane, where we have already seen (Figure \ref{f:EW_vs_b}) that TNG50 is consistent with observations. Therefore, for the remainder of this paper we restrict our observational comparison to sight lines near the major axis.

\begin{figure*}
\gridline{
\fig{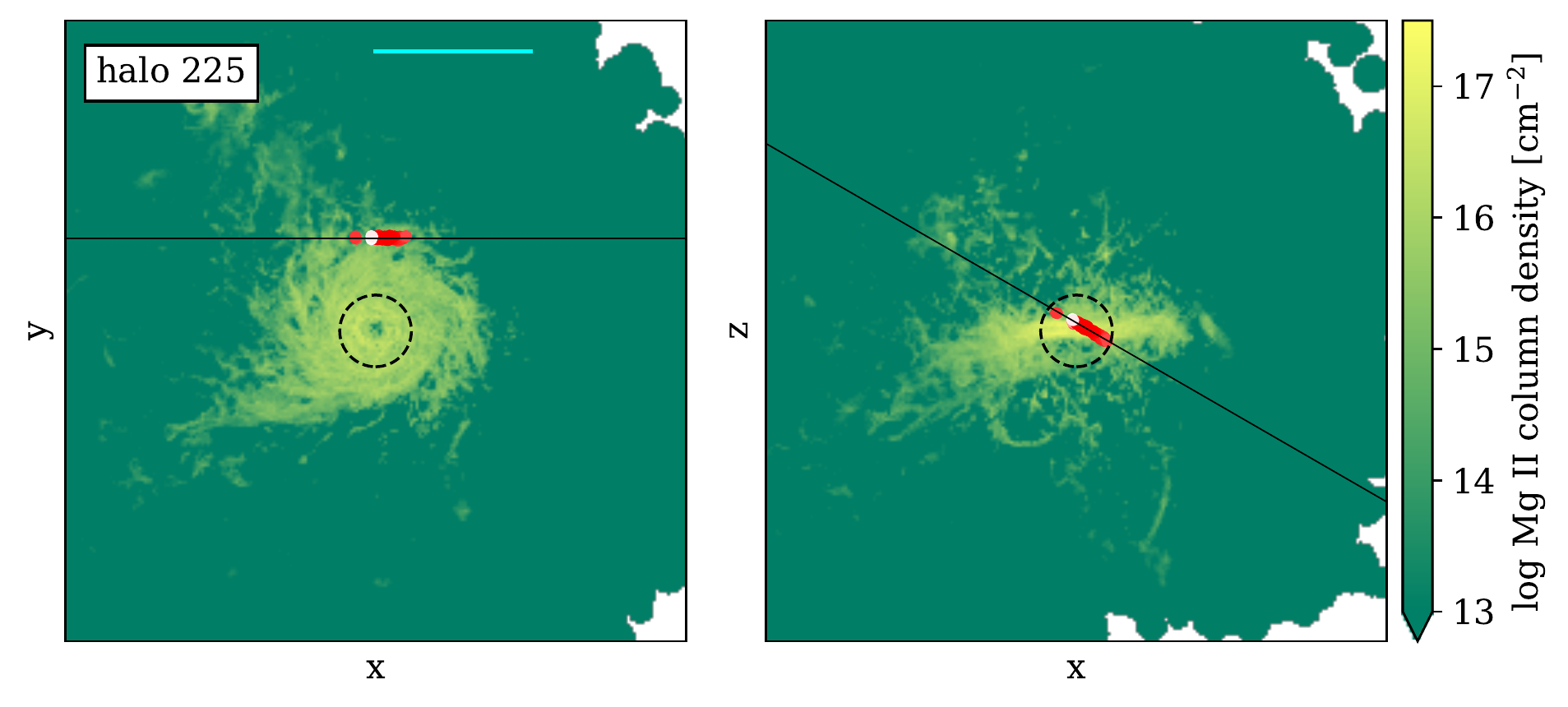}{0.55\textwidth}{}
\fig{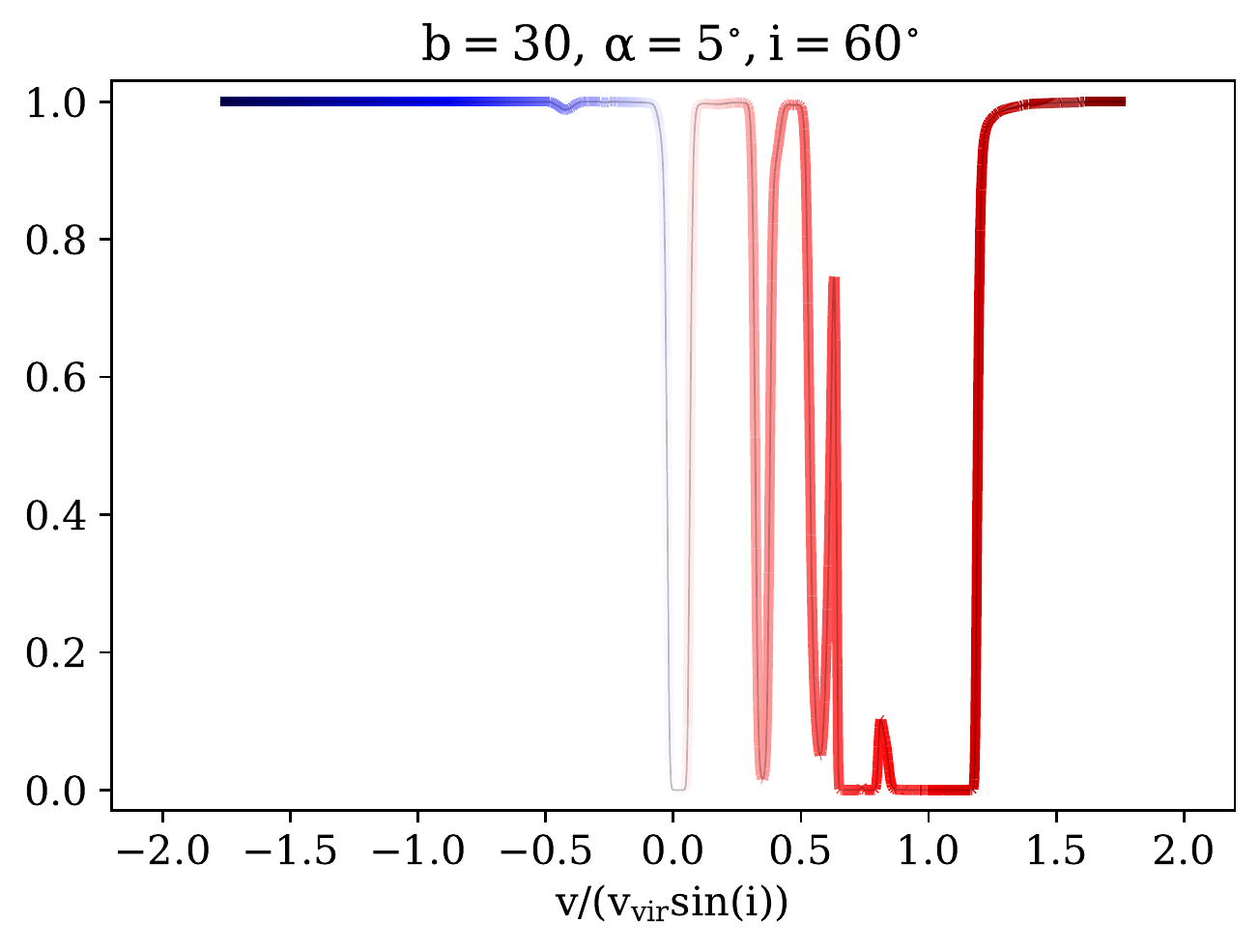}{0.33\textwidth}{}
}
\vspace{-20pt}
\gridline{
\fig{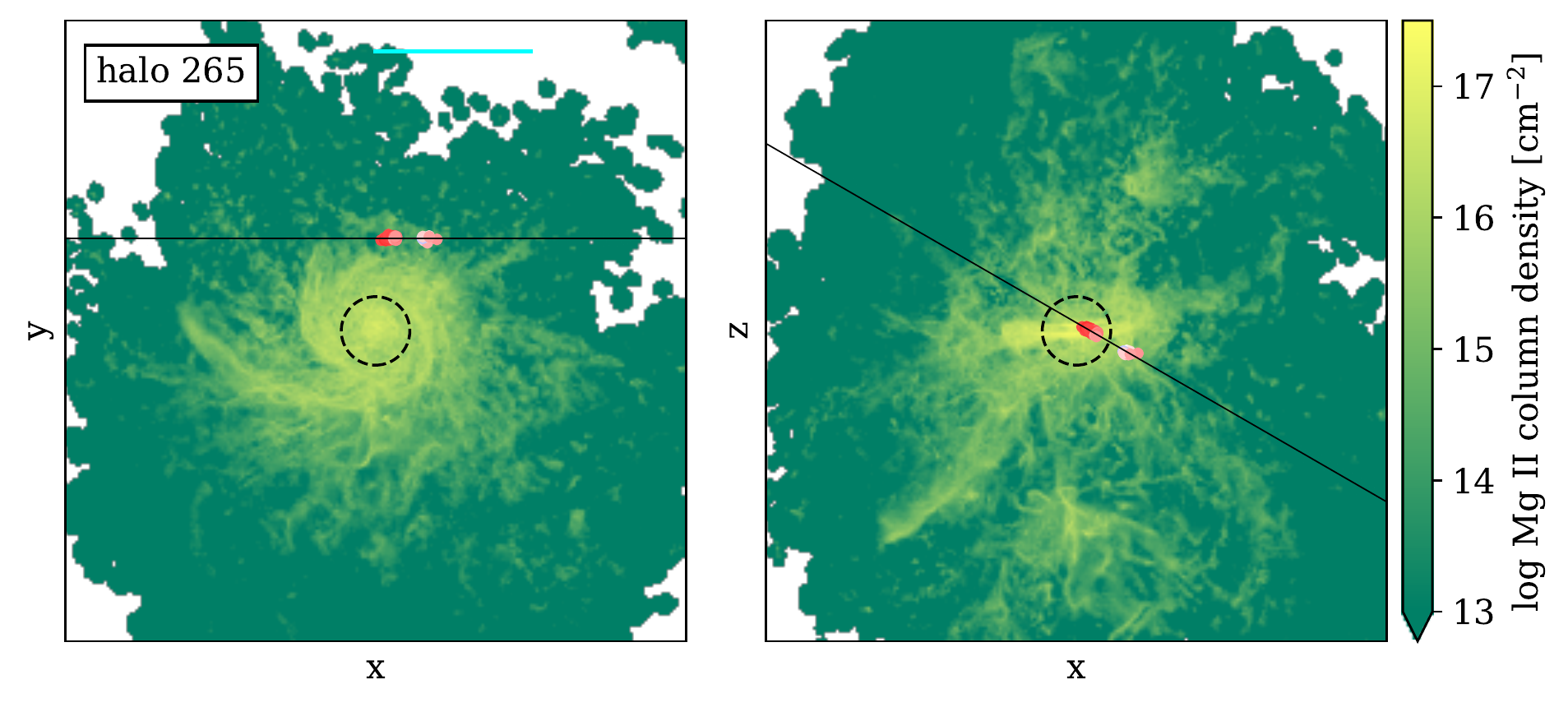}{0.55\textwidth}{}
\fig{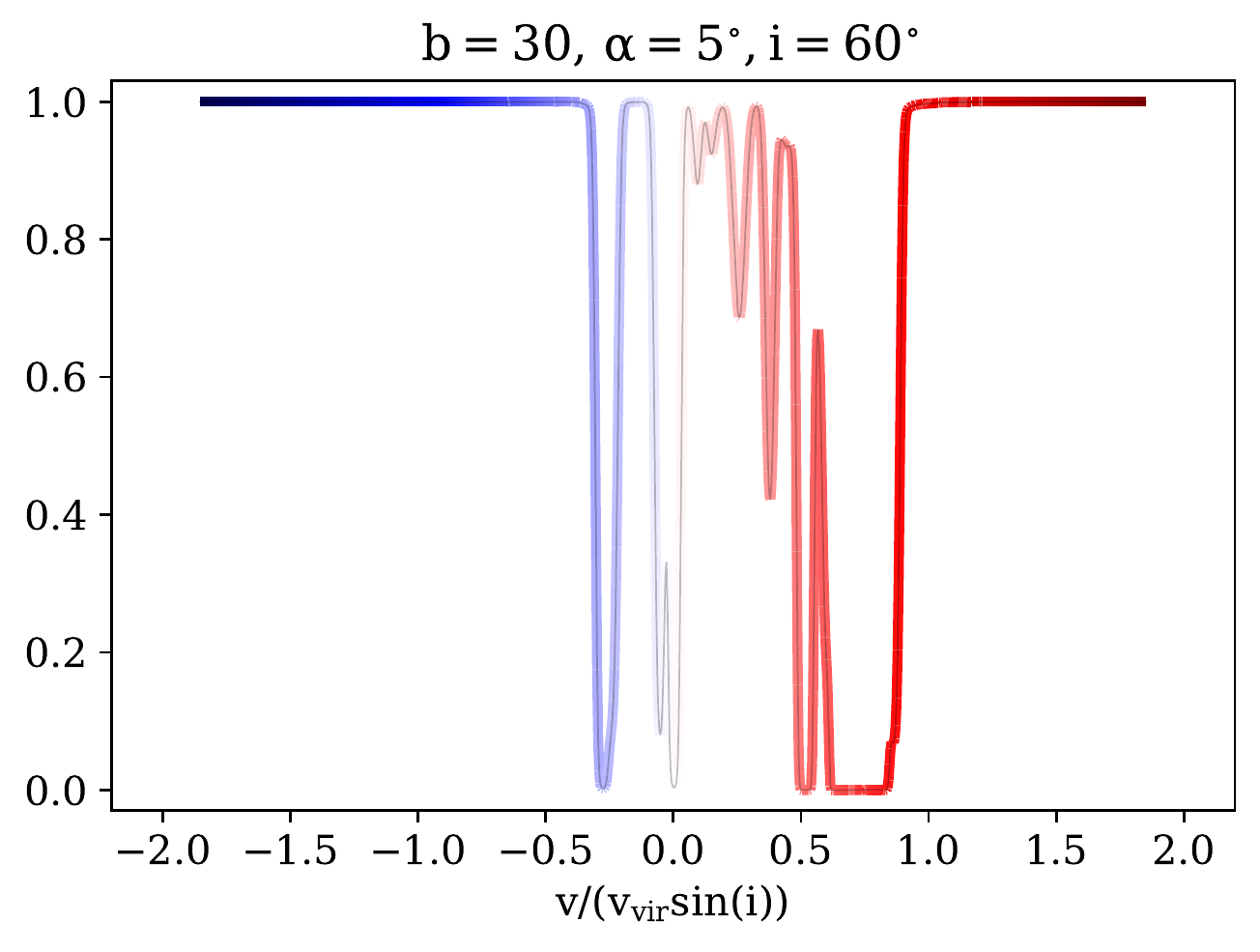}{0.33\textwidth}{}
}
\vspace{-20pt}
\gridline{
\fig{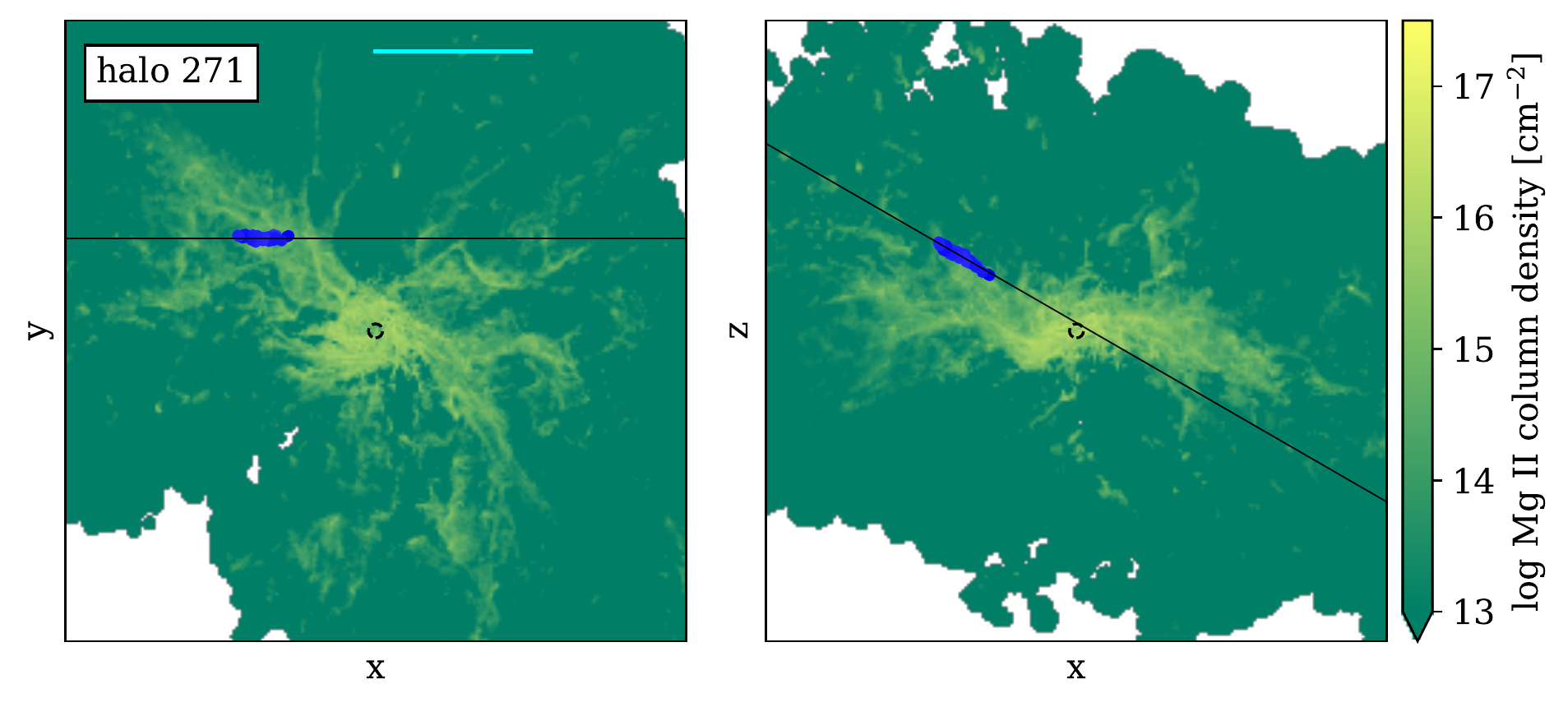}{0.55\textwidth}{}
\fig{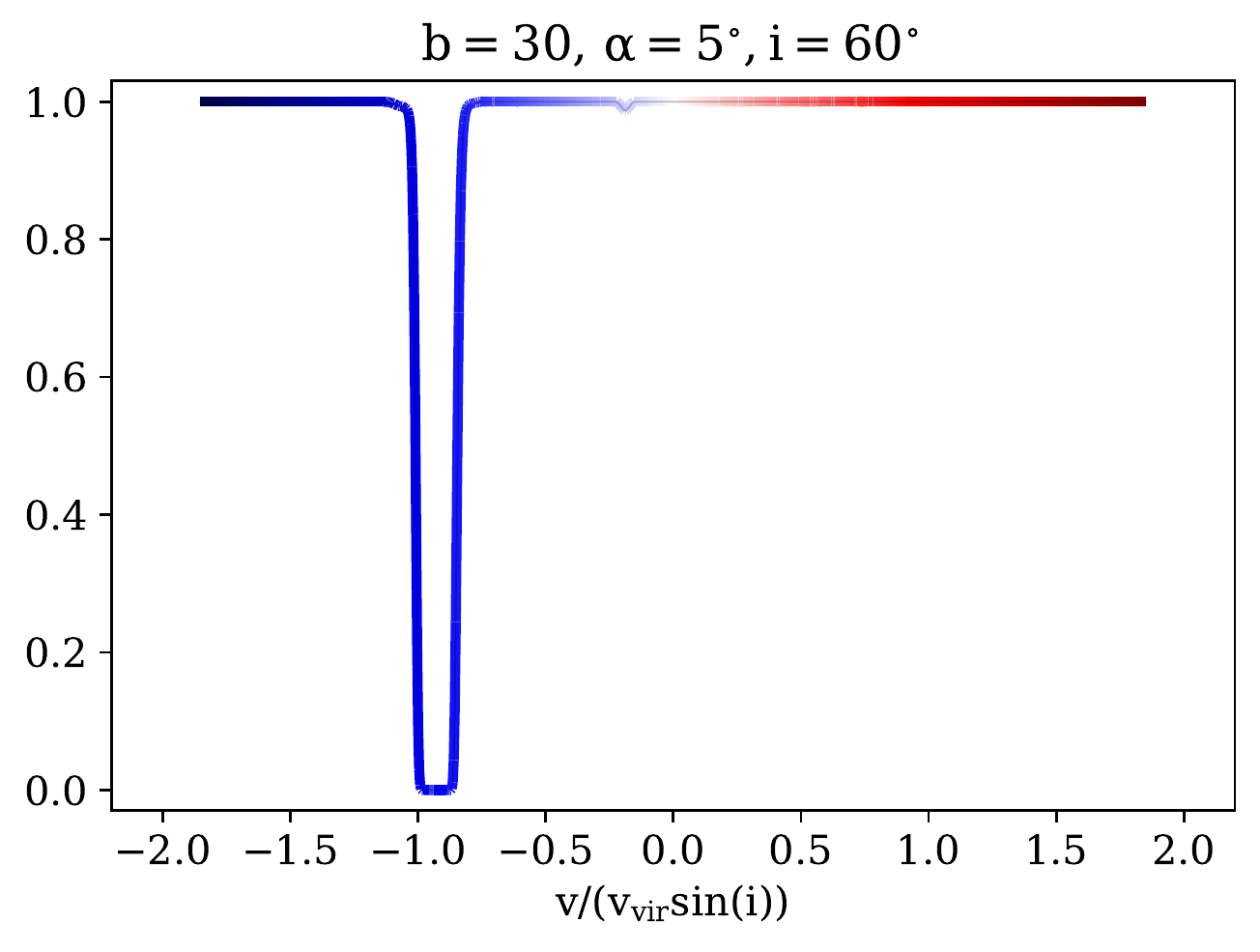}{0.33\textwidth}{}
}
\vspace{-20pt}
\gridline{
\fig{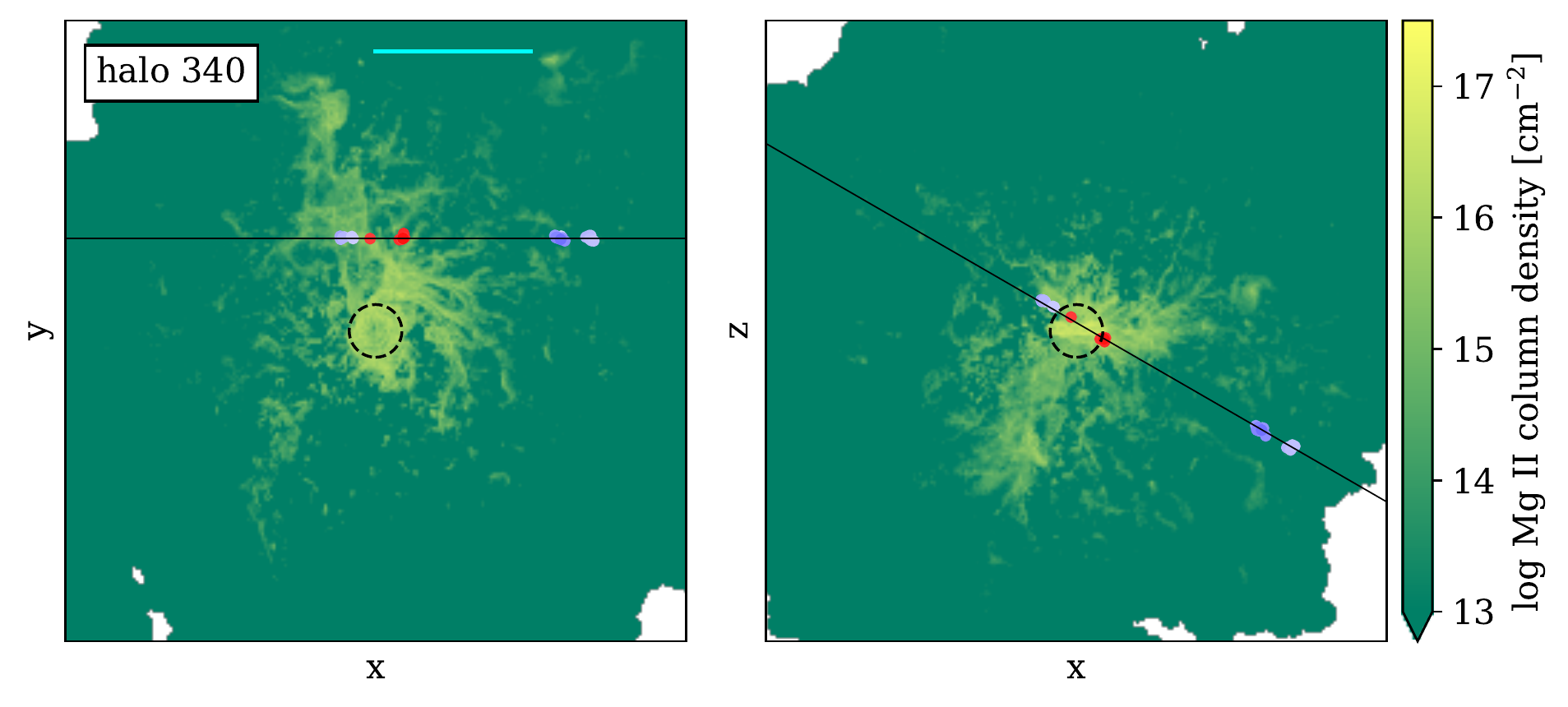}{0.55\textwidth}{}
\fig{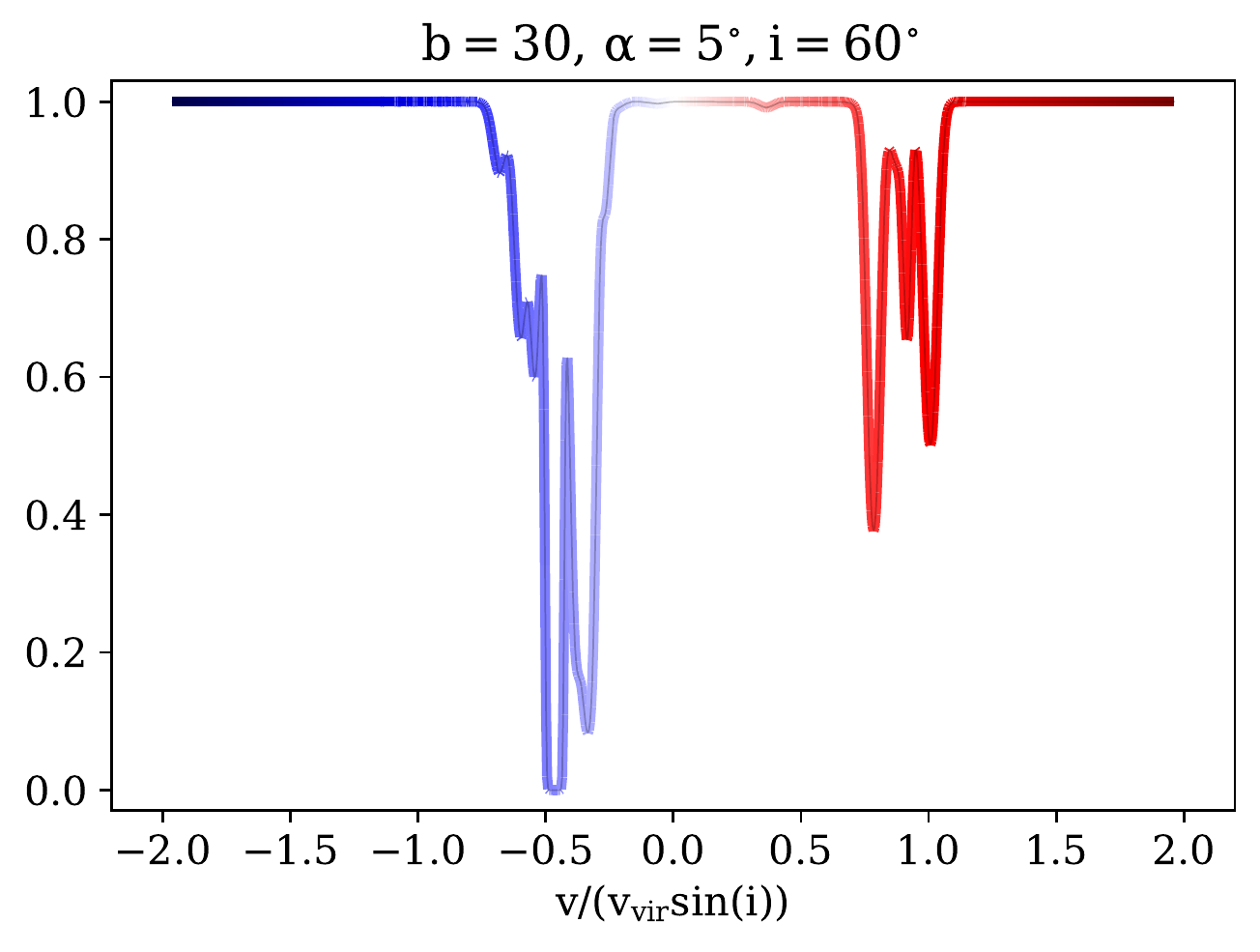}{0.33\textwidth}{}
}
\vspace{-15pt}
\caption{Each row contains two \MgII{} column density maps of a halo from Figure \ref{f:MgII_maps} projected along the vertical (left) and a horizontal (middle) axis. A sight line at $b=30 \; \rm{kpc}$, $\alpha = 5^{\circ}$, and $i=60^{\circ}$ is overlaid along with gas cells that intersect that sight line and have a \MgII{} column density of at least $10^{12} \; \rm{cm^{-2}}$, which accounts for $>95\%$ of the \MgII{} mass along those sight lines. The \MgII{} gas cells and the resulting flux-normalized velocity spectrum (right) are colored by the velocity along the line of sight normalized by $V_{\rm{vir}}\sin(i)$, where $V_{\rm{vir}}$ is the virial velocity of the halo. Dashed circles show twice each galaxy's stellar half-mass--radius. 
}
\label{f:MgII_sightlines}
\end{figure*}

Having established the degree of consistency of \MgII{} equivalent widths, we now examine kinematic signatures of \MgII{} along sight lines in TNG50 and compare them to MEGAFLOW. In Figure \ref{f:MgII_sightlines}, we explicitly draw the connection between the \MgII{} gas cells that contribute to the column densities seen in Figure \ref{f:MgII_maps} and the velocity spectrum created from a subset of those cells that intersect a sight line through the halo. In each row, we show two orientations of one of the four halos from Figure \ref{f:MgII_maps} overlaid with a sight line with $b=30 \; \rm{kpc}$, $\alpha = 5^{\circ}$, and $i = 60^{\circ}$, and the \MgII{} velocity spectrum generated from that sight line. From these few examples it is clear that the gas producing the \MgII{} absorption is generally not distributed uniformly along any sight line: it is usually concentrated in discrete clumps in regions of the sight line nearest to the galaxy. This is seen clearly in rows one, two, and four of Figure \ref{f:MgII_sightlines}, where the majority of gas cells have positive LOS velocities (i.e.,~corotating with the galaxy) and produce distinct kinematic components in the spectrum that are often saturated. 

It is also notable that by comparing the spectra alone it is possible to distinguish morphological differences in the \MgII{} distribution between halos. The first two halos, for example, have a prominent \MgII{} disk that both spectra reveal to be primarily corotating. The halo in row three, however, does not have such a clear disk, and the spectrum is instead composed of a cluster of counter-rotating gas cells significantly above the plane of the galaxy. The halo in row four has a spectrum with substantial corotating and counter-rotating components, which imply \MgII{} structure in between the ordered halos (rows one and two) and disordered ones (row three). With this small sample, we have demonstrated that the velocity spectrum, despite being composed of a very small fraction of all of the \MgII{} gas, is capable of reflecting the potential diversity of \MgII{} gas kinematics in halos of similar mass, but is also fairly consistent between halos with similar morphologies. Later in the paper, we consider whether the \MgII{} gas reflects the kinematics of other components of the CGM.

\begin{figure*}
\fig{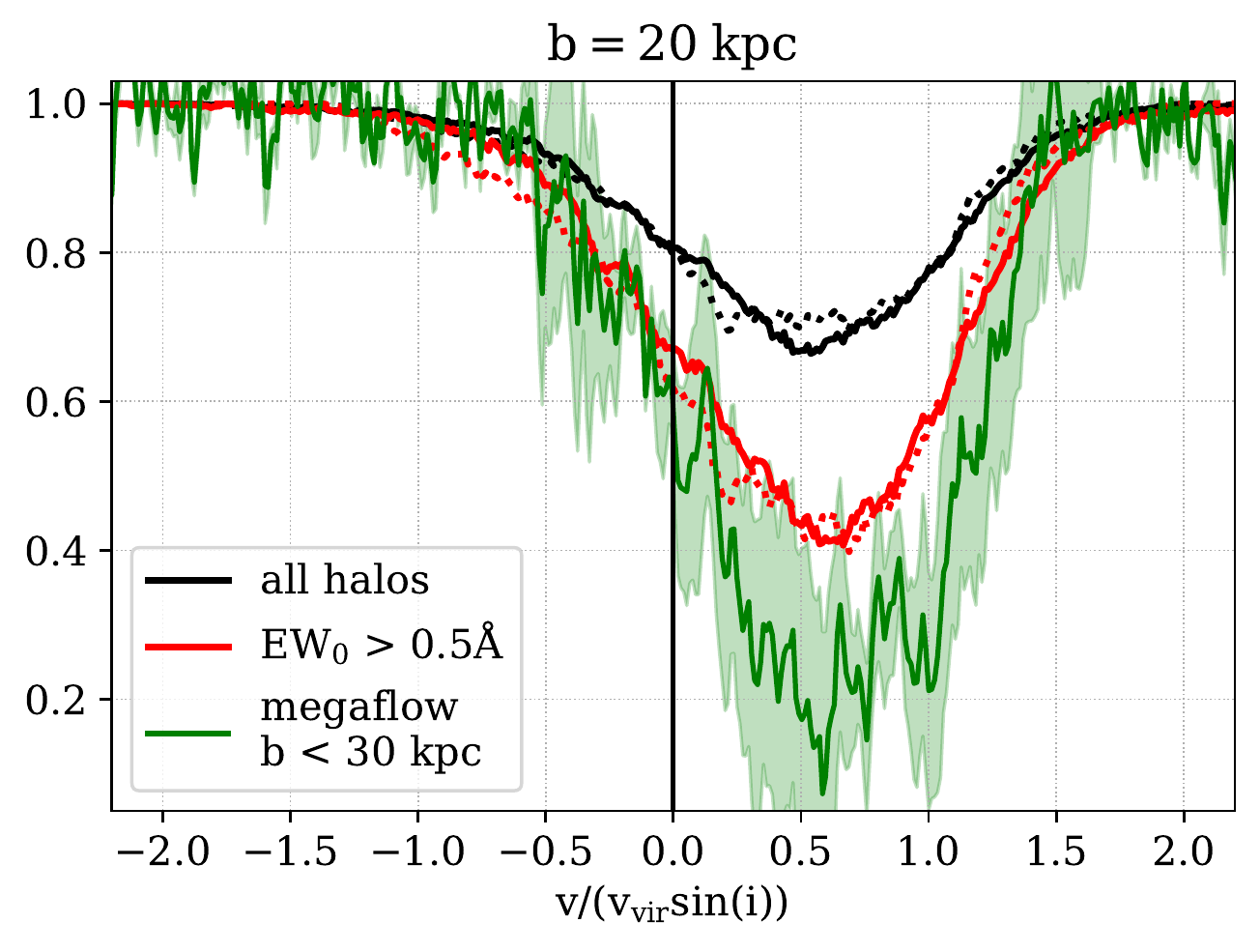}{0.48\textwidth}{}
\fig{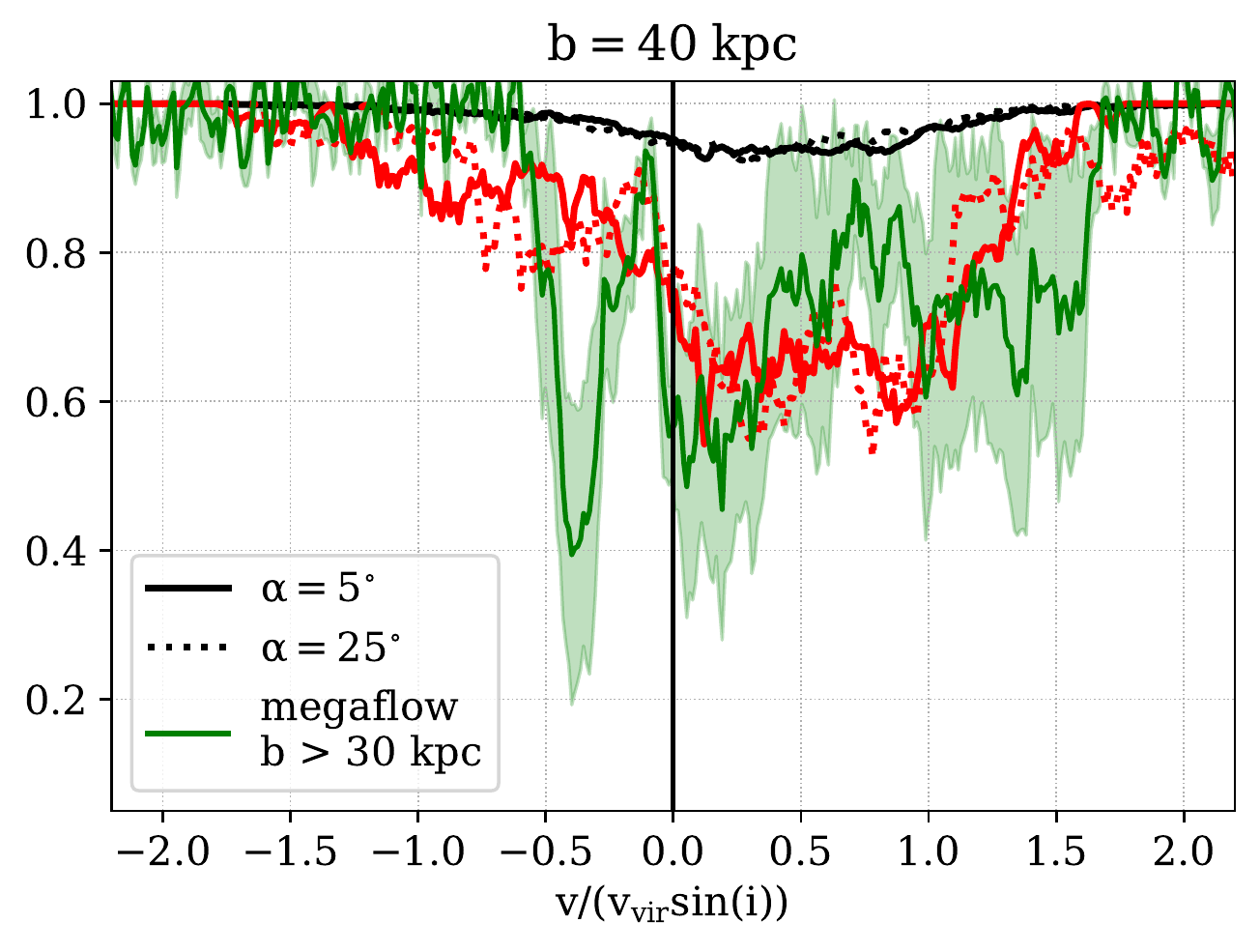}{0.48\textwidth}{}
\vspace{-15pt}
\caption{The stacked \MgII{} velocity spectra for the full fiducial TNG50 sample (black) and the subset of strong absorbers (red) for sight lines with $\alpha = 5^{\circ}$ (solid) and $25^{\circ}$ (dotted), and $b = 20 \; \rm{kpc}$ (left) and $40 \; \rm{kpc}$ (right). Spectra are normalized by $V_{\rm{vir}}\sin(i)$, where $V_{\rm{vir}}$ is the halo's virial velocity. The green line in each panel is the stacked spectrum of the four smallest (left) and largest (right) impact parameters from \citet{Zabl19}, and the green shaded region is an estimate of the error from bootstrapping. 
}
\label{f:spectra}
\end{figure*}

From these results, we now compare stacked spectra from the fiducial sample to the stacked spectra presented in \citet{Zabl19}. Figure~\ref{f:spectra} shows stacked spectra for the entire TNG50 fiducial sample (black), TNG50 strong absorbers (red), and the absorbers from \citet{Zabl19} (green). The two panels correspond to two different impact parameters that allow a comparison between absorbers nearer to a galaxy and farther from a galaxy. In the left panel, showing stacked spectra at small impact parameters, there is a very clear kinematic picture. The strong absorber spectrum from TNG50 is symmetric, centered at $\approx0.6 V_{\rm{vir}}$, and has a with a full width at half maximum (FWHM) of $1.2 V_{\rm{vir}}$, the same as the spectrum of \citet{Zabl19}. Thus, qualitatively, strong \MgII{} absorbers as a population generally have LOS velocities in the same direction as their corresponding galaxies' rotations. One slight difference with the stacked spectra for strong absorbers is that the TNG50 spectrum (red) is somewhat shallower than the observed spectrum (green). However, there is essentially no difference between TNG50 spectra generated from sight lines at the two azimuthal angles $\alpha = 5^{\circ}$ (solid line) and $25^{\circ}$ (dotted line).

In Figure~\ref{f:spectra} (left), the only difference between the full fiducial spectrum and the strong absorber-only spectrum is the depth, indicating that, as a population, strong absorbers are not kinematically distinct from absorbers in general at this impact parameter. The precise reason for the discrepancy in the depth is difficult to determine, but it may be sensitive to certain parameters in the TNG physics model (e.g., metal loading of outflows from supernovae). However, it could also be an effect of simulation resolution (see Section \ref{sec:results3}). So, while TNG50 potentially slightly underproduces the observed amount of \MgII{} gas at 20 kpc, it does possess average kinematics that are consistent with observations of the same region of the CGM. 

Figure~\ref{f:spectra} (right) compares the stacked spectra at a larger impact parameter ($b=40$ kpc). The strong absorbers from TNG50 and MEGAFLOW \citep{Zabl19} are both shallower, wider (FWHMs of $1.3 V_{\rm{vir}}$ and $2 V_{\rm{vir}}$ respectively), no longer symmetric, and significantly noisier, though both are still approximately centered at a velocity on the order of $V_{\rm{vir}}/2$. At this impact parameter, the depths of the simulated strong absorber and observed spectra are consistent with each other. However, strong absorbers no longer kinematically resemble the full fiducial sample: in addition to being much rarer at 40 kpc than at 20 kpc, the strong absorbers have larger positive velocities, indicating that \MgII{} in this region is tracing atypically faster-moving gas. As was the case at 20 kpc, the difference in the spectra between the two azimuth angles is minor. We also note here, but do not show, that the shapes and depths of individual  spectra from \citet{Zabl19} match quite well with particular individual spectra from the much larger fiducial sample from TNG50 (examples of individual spectra from TNG50 are shown in Figure~\ref{f:MgII_sightlines}). 

\subsection{3D Kinematics of \MgII{} in TNG50}
\label{sec:results2}

\begin{figure}
\fig{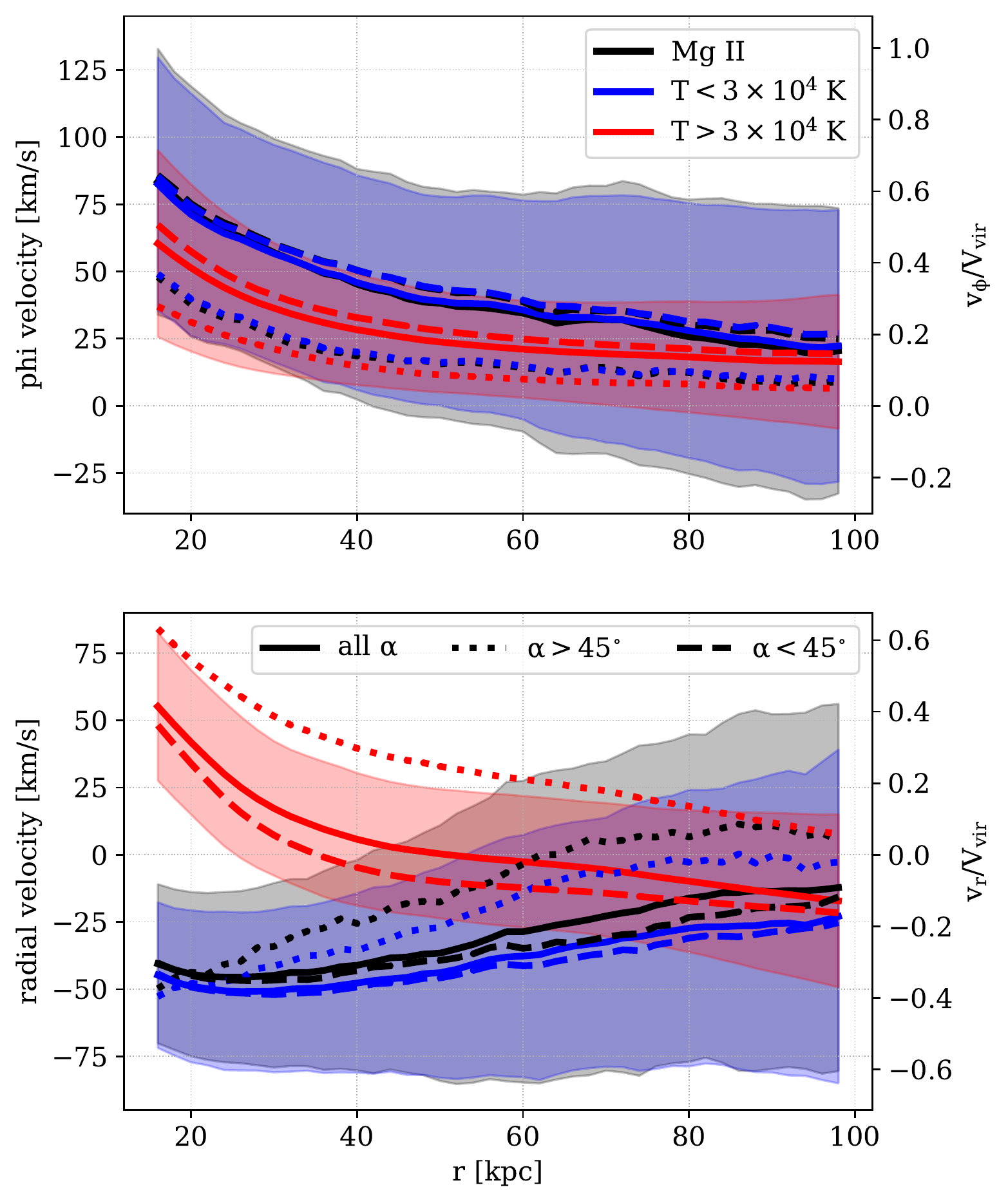}{0.48\textwidth}{}
\vspace{-25pt}
\caption{Mean mass-weighted velocity profiles of the spherical phi-component ($v_{\phi}$, top) and r-component ($v_{r}$, bottom) for cold gas (blue), hot gas (red), and \MgII{} gas (black) in spherical bins. Velocity is given in $\rm{km \; s^{-1}}$ and as a fraction of the virial velocity. A temperature of $3\times10^4 \; \rm{K}$ is used to separate ``cold'' and ``hot'' gas. Profiles are shown for gas in the entire halo (solid), gas with $\alpha > 45^{\circ}$ (dotted), and gas with $\alpha < 45^{\circ}$ (dashed). Shaded regions show the $\pm1\sigma$ scatter of the solid lines and are of similar size for all profiles.
}
\label{f:vprofiles}
\end{figure}

In this section, we characterize the three-dimensional kinematics of the \MgII{} gas in TNG50 and its relation to the observed quantities we discussed in Section~\ref{sec:results1}. We show average velocity profiles of the halos in the fiducial sample in Figure~\ref{f:vprofiles}. The top panel shows the azimuthal velocity component ($v_{\phi}$) in spherical coordinates as a function of radius. We divide gas into cold and hot components based on a temperature threshold of $3\times10^4$ K, which is chosen to separate the cold and hot clusters seen in Figure~\ref{f:MgII_phase}, although the profiles are not sensitive to the precise choice of temperature threshold. To understand the relationship of the hot and cold gas to \MgII-bearing material we also show the \MgII{} mass-weighted profiles.

First, we see that the \MgII{} gas and the cold gas have nearly identical $v_{\phi}$ profiles throughout the halo. In the innermost regions of the CGM ($15-20 \; \rm{kpc})$, the cold gas has a mean azimuthal velocity of $80 \; \rm{km} \; \rm{s^{-1}}$ ($\approx 0.6 V_{\rm{vir}}$), while in the outermost regions ($90-100 \; \rm{kpc})$, the mean azimuthal velocity decreases to $20 \; \rm{km} \; \rm{s^{-1}}$ ($\approx 0.15 V_{\rm{vir}}$). At all radii, the $\pm1\sigma$ scatter is quite large ($\approx 100 \; \rm{km} \; \rm{s^{-1}}$), though the standard errors on this and all other mean velocities in Figure \ref{f:vprofiles} range from only $1-3 \; \rm{km} \; \rm{s^{-1}}$. Though not explicitly shown, most of the cold and \MgII{} gas mass is closer to the major rather than the minor axis because the all-$\alpha$ profiles are much more similar to the $\alpha < 45^{\circ}$ (dashed) profiles than the $\alpha > 45^{\circ}$ (dotted) profiles. Hot gas has lower azimuthal velocities at all radii, a slightly shallower slope to its profile, and a smaller scatter in azimuthal velocity by a factor of $\approx 2$ but is otherwise qualitatively similar to the cold and \MgII{} gas. This relationship between hot and cold gas is consistent with similar measurements of $v_{\phi}$ made from TNG100 in \cite{DeFelippis20}. 

In the radial-velocity profiles (Figure \ref{f:vprofiles}, bottom), we see a gulf between the velocities of the hot and cold gas develop within $90 \; \rm{kpc}$. Above this radius, the average radial velocities of all components of the gas converge to $-20 \; \rm{km} \; \rm{s^{-1}}$ ($\approx 0.15 V_{\rm{vir}}$), though the spread of radial velocities in this region of the CGM is very large, especially for cold gas ($\pm1\sigma$ scatter of $120 \; \rm{km} \; \rm{s^{-1}}$). Moving toward smaller radii, the cold gas inflow velocities become larger, while hot gas inflow velocities decrease and then switch to a net outflow at $50 \; \rm{kpc}$. The \MgII{} gas still traces the cold gas, which reaches typical inflowing velocities of $45-50 \; \rm{km} \; \rm{s^{-1}}$ in the inner CGM out to $r=40 \; \rm{kpc}$, where the spread in radial velocities is a factor of $2$ smaller than in the outer halo. The geometry of accretion and outflows is evident from this panel as well: hot gas has especially large mean outflowing velocities for $\alpha > 45^{\circ}$ while cold gas in the same region has a mean inflowing velocity in the inner halo and nearly no net radial motion in the outer halo. Most of the cold and \MgII{} gas mass is moving toward the galaxy in regions surrounding the major axis out to a substantial fraction of the virial radius. It is also clear that kinematically, \MgII{} gas in TNG50 is nearly identical to a simple cut on temperature and so is an excellent tracer of the kinematics of cold CGM gas. In the context of Section \ref{sec:results1}, these results indicate that mock \MgII{} spectra are representative of the entire cold phase of the CGM along the same sight lines. 

\begin{figure}
\fig{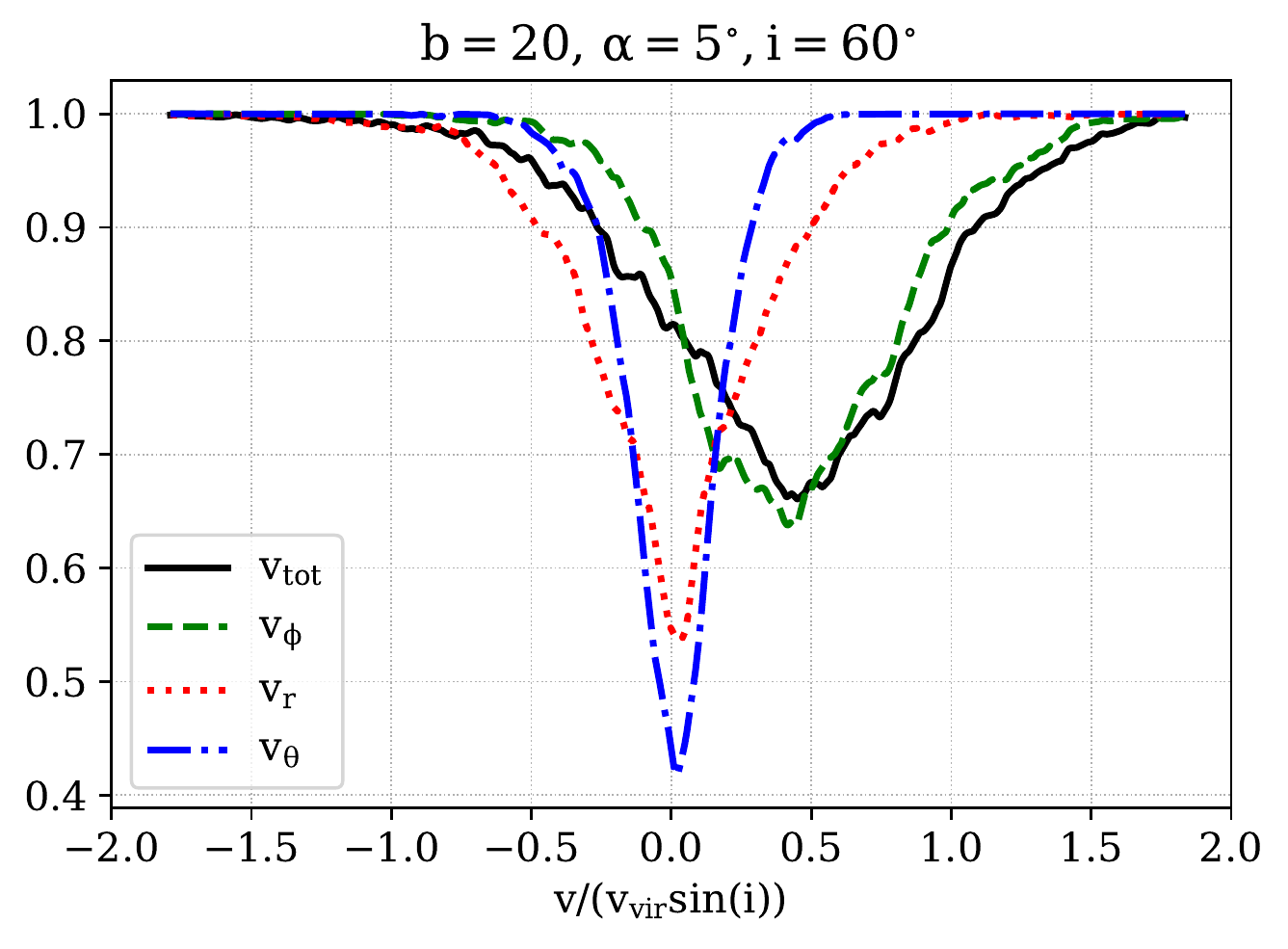}{0.48\textwidth}{}
\vspace{-25pt}
\caption{Stacked \MgII{} velocity spectra for the full fiducial TNG50 sample at a single sight line. The contributions of the three spherical components of velocity -- $v_{r}$ (dotted red), $v_{\phi}$ (dashed green), and $v_{\theta}$ (dotted-dashed blue) -- are shown, as well as the spectrum created from the total velocity (solid black).}
\label{f:vcomponents}
\end{figure}

Finally, we examine the 3D velocities of the \MgII{} gas along our sight lines. In Figure \ref{f:vcomponents}, we plot stacked spectra for \MgII{} using the three spherical velocity components individually ($r$, $\theta$, and $\phi$), and compare those to the spectrum generated with the full velocity of our fiducial sample of halos. Both the $r$ and $\theta$ component spectra are centered at $0 \; \rm{km} \; \rm{s^{-1}}$, indicating that over the entire sample they do not contribute any net velocity shift to the gas along the sight lines. The spectrum of the $\phi$ component is remarkably similar to the spectrum of the entire velocity, both in terms of velocity shift and width. This means that for our fiducial sample, the shape of the stacked velocity spectrum along sight lines is completely determined by only the $\phi$ (i.e., rotational) component of the velocity along those sight lines. 

\subsection{Effects of halo mass and resolution on \MgII{} in TNG50}
\label{sec:results3}

\begin{figure*}
\fig{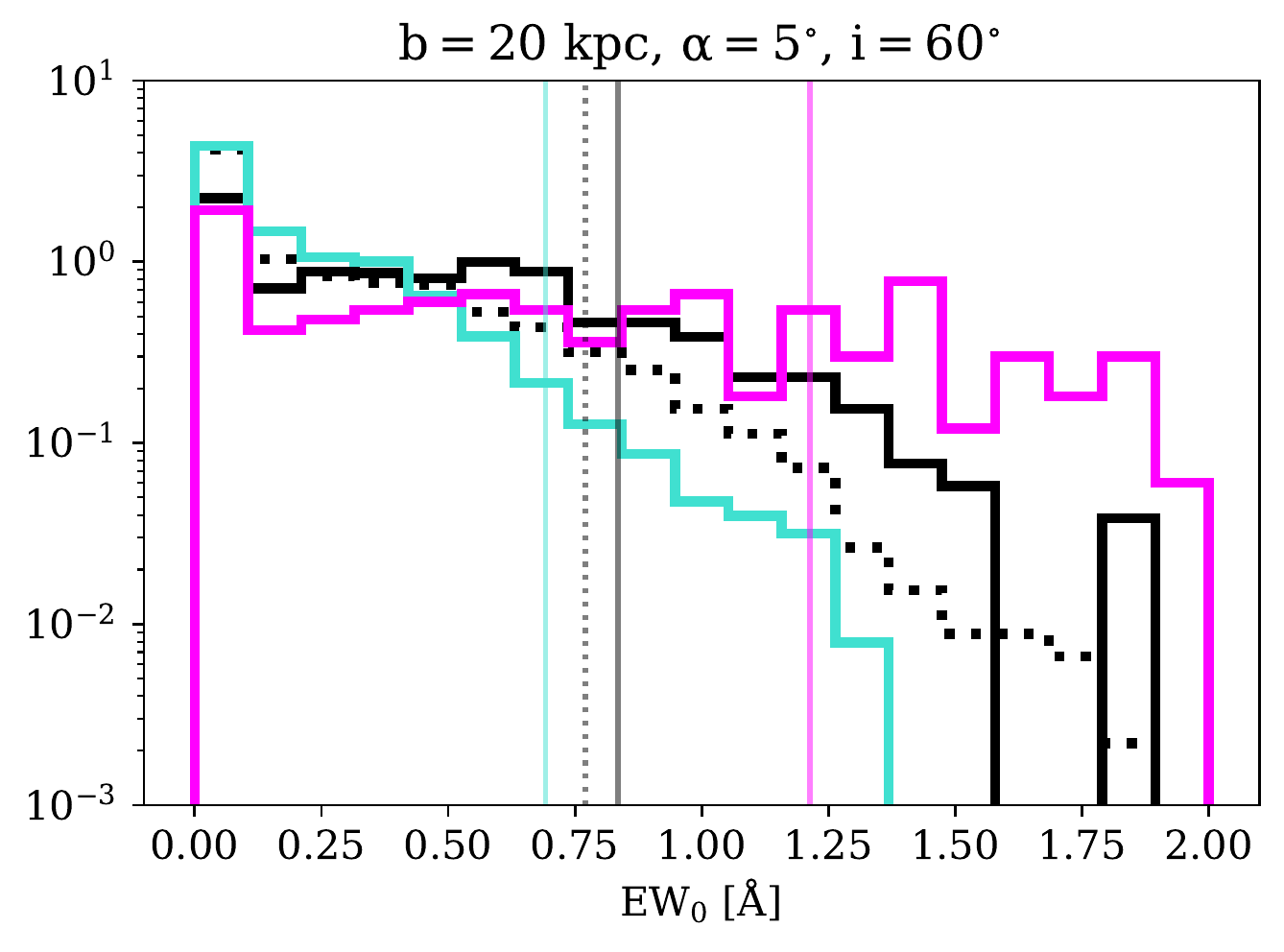}{0.48\textwidth}{}
\fig{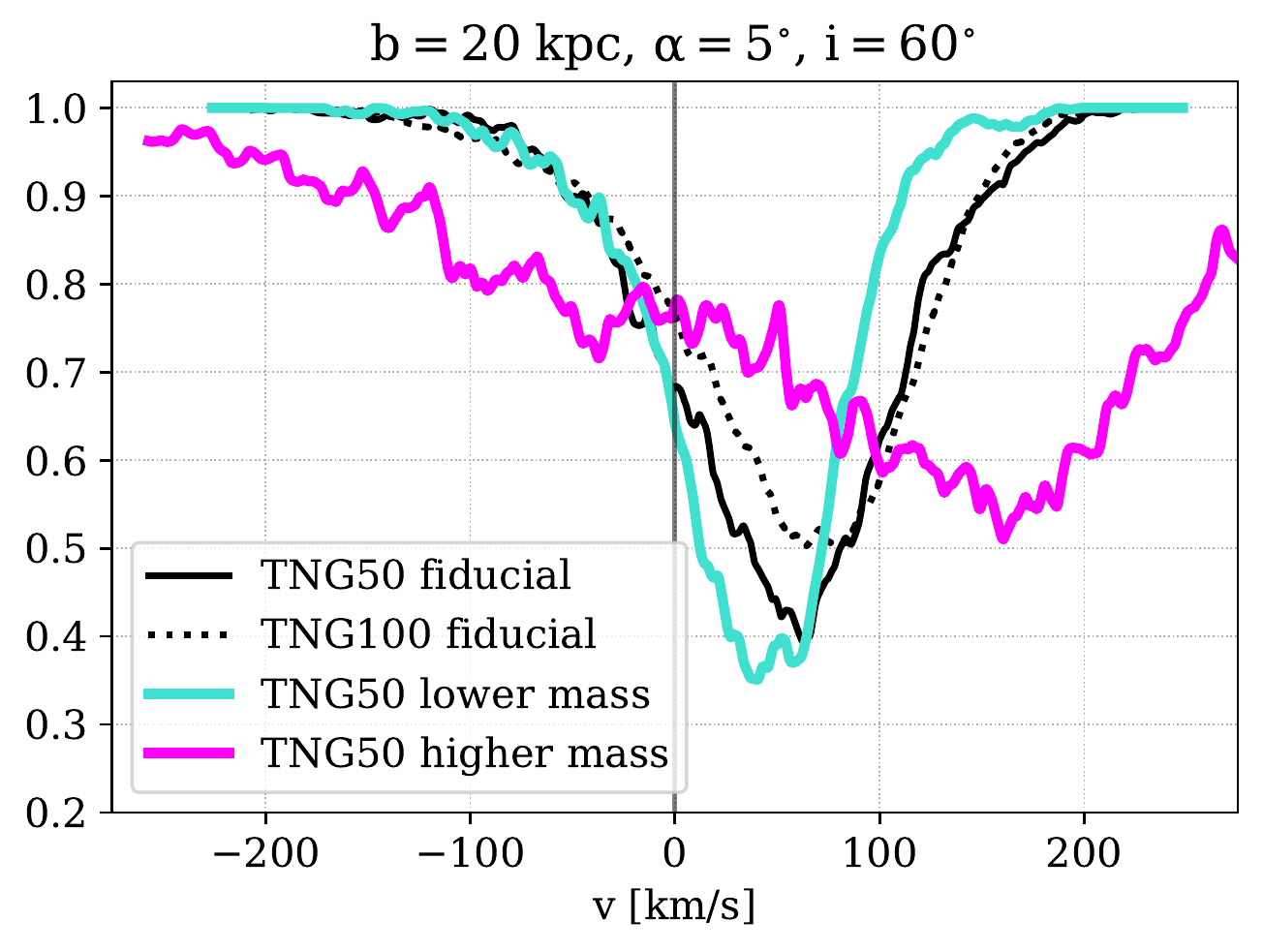}{0.48\textwidth}{}
\vspace{-15pt}
\caption{Left: rest-frame equivalent-width distribution of the TNG50 fiducial sample (solid black), lower-mass halos with $10^{11} \; M_{\odot} < M_{\rm{halo}} < 10^{11.5} \; M_{\odot}$ (cyan), higher-mass halos with $10^{12} \; M_{\odot} < M_{\rm{halo}} < 10^{12.5} \; M_{\odot}$ (magenta), and the same mass halos from TNG100 (dotted black) at the same sight line of $b=20 \; \rm{kpc}$, $\alpha = 5^{\circ}$, and $i=60^{\circ}$. The mean EW$_0$ of the strong absorbers in each halo mass bin is shown with a translucent vertical line of the same color. Right: stacked velocity spectra of the same halo samples with velocities in $\rm{km \; s^{-1}}$.}
\label{f:mass_and_resolution}
\end{figure*}

We now describe how our main results vary with halo mass and mass resolution. To study the effect of halo mass, we consider two mass bins containing halos from TNG50 with $10^{11} \; M_{\odot} < M_{\rm{halo}} < 10^{11.5} \; M_{\odot}$ and $10^{12} \; M_{\odot} < M_{\rm{halo}} < 10^{12.5} \; M_{\odot}$  at $z=1$, which are above and below the fiducial mass range and contain 1130 and 167 halos, respectively. As in Section \ref{sec:results1} we calculate \MgII{} equivalent widths and generate velocity spectra that we show in Figure \ref{f:mass_and_resolution}. For easier comparison, we also show the TNG50 fiducial sample. 

As shown in the left panel of Figure \ref{f:mass_and_resolution}, at a given impact parameter, the shape of the equivalent-width distribution changes with halo mass: lower halo masses (cyan) are much more likely to host weak or nonabsorbers than higher halo masses (magenta), and they are much less likely to host strong absorbers. We find this trend to hold at all impact parameters studied in this paper. We can see the effect on observability with the vertical lines in this panel, which show the mean equivalent widths of the strong absorbers in each mass bin. Typical strong absorbers in the fiducial sample have only slightly larger equivalent widths than those those at lower halo masses, but are substantially weaker than the strong absorbers at higher halo masses. At larger impact parameters, the mean equivalent widths of all strong absorbers is $\approx 0.8 \; \rm{\AA}$, but they are exceedingly rare in lower-mass halos. Thus, the primary effects of increasing halo mass on strong absorbers are to increase their occurrence at all impact parameters, especially at large distances, and to increase the mean equivalent width of strong absorbers for halo masses $\gtrsim 10^{12} \; M_{\odot}$. We note that this result is qualitatively consistent with \cite{Chen10b}, who find a larger \MgII{} extent in the CGM of higher-mass galaxies.

Also shown in the left panel of Figure \ref{f:mass_and_resolution} is the equivalent-width distribution of 4315 halos with the same mass as the fiducial sample from the TNG100 simulation, which has a lower baryonic mass resolution than TNG50 by a factor of $\sim 16$. Decreasing the simulation resolution lowers equivalent widths overall and steepens the distribution in the same way as decreasing the halo mass does, but the effect is weaker. The mean equivalent width of strong absorbers is largely unaffected by the change in resolution. 

In the right panel of Figure \ref{f:mass_and_resolution} we examine the effect of halo mass and resolution on the observed \MgII{} spectrum of strong absorbers. We note that the spectra of the entire samples, as in Figure \ref{f:spectra}, have the same shape and center as their corresponding strong absorber subset, but are substantially shallower. We also plot the real velocity rather than the normalized velocity to emphasize the difference in equivalent widths, which can be more easily read off. 

We see that the fiducial and lower-mass bins have remarkably similar spectra: they are both symmetric and centered at moderate positive velocities. The spectrum of the higher-mass bin is markedly different: it is much broader, asymmetric, and centered at a significantly higher velocity. It still, however, shows a preference for \MgII{} gas to be corotating. We note that the difference between Figure \ref{f:mass_and_resolution} as shown and the corresponding velocity-normalized spectrum (not shown) is that the normalized higher-mass spectrum is compressed slightly and therefore appears more similar to the normalized fiducial spectrum. Additionally, while the lower-mass and fiducial spectra are both centered at $\approx 0.5 V_{\rm{vir}}$, the higher-mass spectra are peaked at $\approx V_{\rm{vir}}$. Higher halo masses ($\gtrsim 10^{12} \; M_{\odot}$) thus have substantially more \MgII{} absorption and more complex kinematic signatures than for the halo masses of the fiducial sample and lower.

Finally, we consider the difference that resolution makes in the \MgII{} absorption spectrum. As was the case with equivalent widths, the difference caused by resolution is smaller than the difference caused by either increasing or decreasing the halo mass. Apart from a slight change in the depth of the spectrum, the kinematic properties of strong absorbers in TNG are essentially resolution independent (see solid vs. dotted curves in Figure \ref{f:mass_and_resolution} for TNG50 and TNG100, respectively). The effect of increasing the resolution of the simulation is therefore primarily to increase the occurrence of strong absorbers at a given halo mass.

\section{Discussion} \label{sec:discussion}

\subsection{The Role of \MgII{} in TNG}

We consider here the ramifications of the detailed analysis of \MgII{} in TNG from Section \ref{sec:results}. In Figure \ref{f:vprofiles}, we found that \MgII{} gas is very well approximated by a simple temperature cut. Therefore, we expect the angular momentum of cold gas in the CGM of TNG galaxies should be very similar to that of \MgII. \cite{DeFelippis20} found cold CGM gas in halos of this mass range and redshift to have higher angular momentum when surrounding high-angular-momentum galaxies, meaning \MgII{} is likely tracing high-angular-momentum gas in the CGM of these halos. As the velocity spectrum's center and shape is almost completely set by the rotational velocity component (see Figure \ref{f:vcomponents}), it should therefore be possible to use \MgII{} velocity spectra from sight lines near the major axis to estimate the angular momentum of cold gas in the CGM. 

In Section \ref{sec:results3} we examined possible halo mass and resolution dependencies of our results with two main goals in mind: to establish any broad effects of the TNG feedback model on \MgII, and to determine to what extent the cosmological simulation can capture \MgII{} kinematics. Feedback is known to be important for regulating gas flows into, out of, and around galaxies, and therefore could have observable signatures in the \MgII{} spectra, especially at different halo masses. The results of the halo mass analysis suggest that for halos with masses between $10^{11} \; M_{\odot}$ and $10^{12} \; M_{\odot}$, the physical mechanisms affecting their CGM are similar enough to result in \MgII{} spectra that essentially scale with the halo's virial velocity. This is presumably because feedback from supernovae is the dominant form of feedback that affects the CGM for all halo masses below $\sim10^{12} \; M_{\odot}$ and produces \MgII{} gas with similar kinematic signatures. For halos above $10^{12} \; M_{\odot}$ however, \MgII{} gas has stronger overall absorption, as reflected by their flatter EW distribution, and substantially larger velocities and velocity dispersions, as reflected by their very broad velocity spectra. This is likely due to the dominant form of feedback switching from stars to AGN around this halo mass. However, within the higher-mass sample, halos with larger black hole masses do not themselves have broader \MgII{} spectra, so there is probably a combination of effects that result in a noticeable difference in the properties of the spectrum at higher masses. 

\cite{Nelson20} have recently used TNG50 to study the origin of cold \MgII{} gas in the CGM of very massive ($M_{*} \gtrsim 10^{11} \; M_{\odot}$) galaxies and found structures of size a few $\times 10^2 \; \rm{pc}$ that are sufficient to explain the observed covering fractions and LOS kinematics. They also note that while some fundamental properties like the number of cold gas clouds present in halos are not converged at TNG50's resolution, the total cold gas mass of such halos is converged in TNG50. This supports our findings that our kinematic results do not qualitatively change even going from TNG50 to TNG100, a factor of $\sim16$ in mass resolution (Figure \ref{f:mass_and_resolution}), because the majority of the \MgII{} mass is already in the halo by TNG50's resolution. We expect higher-resolution simulations to produce more strong absorbers at a given halo mass but the rotation of \MgII{} near the major axis appears to be a resolution-independent aspect of the CGM for MEGAFLOW analogs in the TNG simulations. 

\begin{figure}
\fig{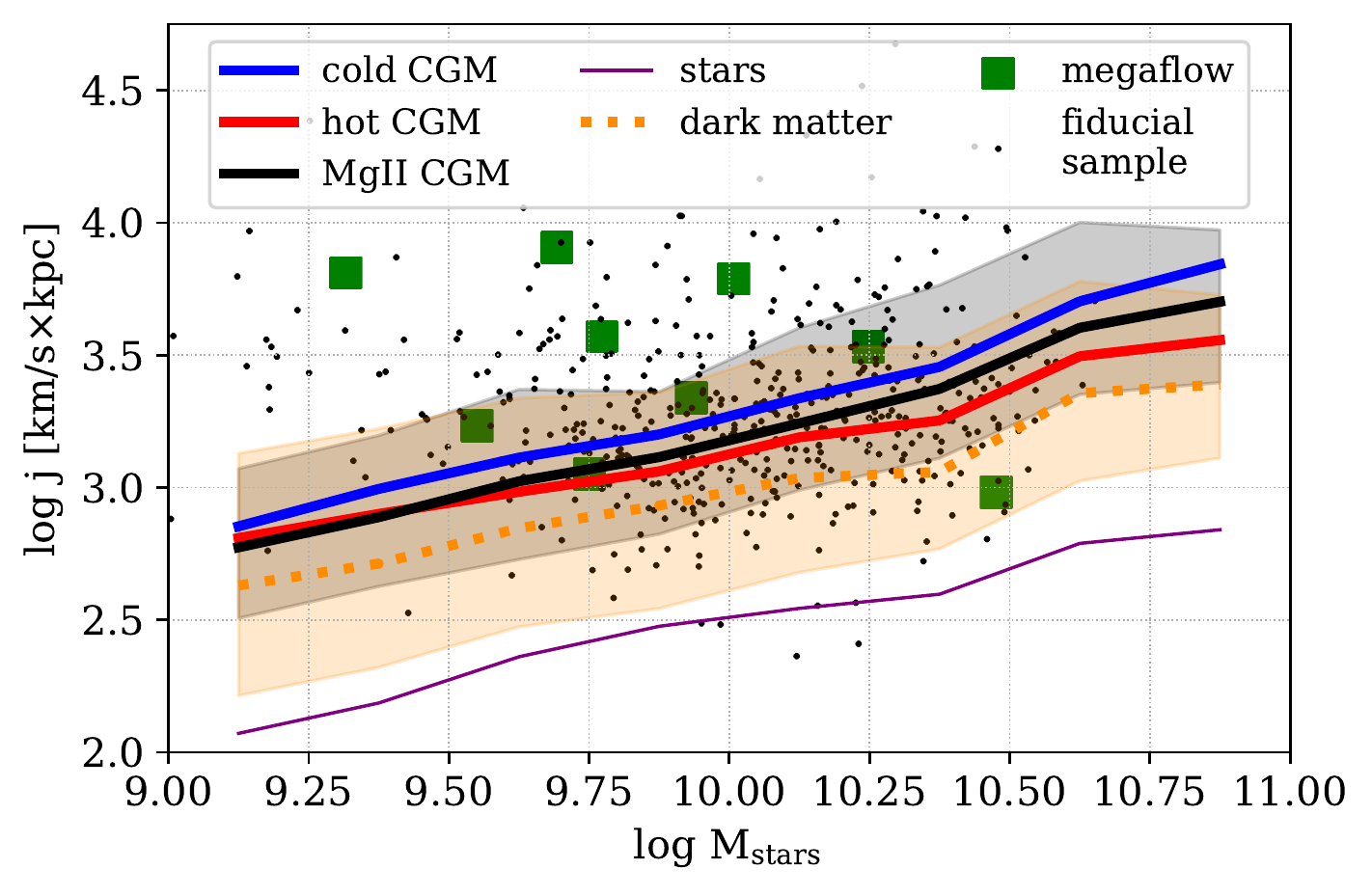}{0.48\textwidth}{}
\vspace{-25pt}
\caption{Median-specific angular momentum vs. galactic stellar mass for the cold (blue), hot (red), and \MgII{} (black) CGM as defined in Figure \ref{f:vprofiles}, as well as the dark matter halo (dotted orange) and the stellar component of the galaxy (purple) at $z=1$. Unlike previous figures, medians are calculated using a sample of all halos containing central galaxies with stellar masses $10^{9} \; M_{\odot} < M_{*} < 10^{11} \; M_{\odot}$. Shaded regions show the 16th and 84th percentiles of the distributions of the \MgII{} gas (black), which is similar in size to all components except dark matter (orange), which has noticeably larger scatter. Black points show the \MgII{} specific angular momentum of the halo-mass-selected fiducial sample that is biased toward higher $j$ for $M_{*} \lesssim 10^{9.75}$. Green squares show estimations for the specific angular momentum of the major-axis absorbers using inferred rotational velocities from \cite{Zabl19}.}
\label{f:jM}
\end{figure}

Finally, in Figure \ref{f:jM}, we show the specific angular momentum ($j$) of different halo components as a function of stellar mass of their central galaxies, with the goal of contextualizing the angular momentum of \MgII{} gas (black line) in the CGM in relation to the rest of the gas in the CGM as well as to the other components of the halo. The slope of this $j-M_{*}$ relation for the stellar component of galaxies (purple line) is $\sim 0.6$ as generally observed \citep[e.g.,][]{Fall13}, and all other components appear to have roughly equal slopes. Most interesting are the relative positions of the CGM and dark matter (orange line) on this plane. At a given stellar mass, all components of the CGM have a slightly higher typical $j$ than that of the dark matter by $\sim 0.2 \; \rm{dex}$. There are multiple potential reasons for this. First, galaxies can remove low-angular-momentum gas from the CGM by accreting it and using it to form stars. Second, feedback from stars and/or AGN can also eject low-angular-momentum gas from the halo completely. Finally, dark matter in the halo can transfer some of its angular momentum to the gas. Regardless, it is clear that \MgII{} traces the angular momentum of the both the cold and hot components of the CGM quite well. 

Also shown in Figure \ref{f:jM} are two sets of points representing \MgII{} gas in individual halos: the fiducial sample in black and the \citet{Zabl19} sample in green, for which $j$ was estimated using their derived rotational velocities. The two are not directly comparable since the points from \citet{Zabl19} represent \MgII{} gas along a single sight line, yet they are still able to reproduce the scatter in this relation found in TNG50, though they are somewhat biased toward higher $j$. This bias is likely due to the selection in \citet{Zabl19} of strong \MgII{} absorption near the major axis, which is where high-$j$ cold gas tends to reside in the CGM as shown in \citet{DeFelippis20}. Nevertheless, from Figure \ref{f:jM} we can conclude that estimations of the angular momentum content of the CGM provided by single sight lines of \MgII{} can get within $\sim0.5 \; \rm{dex}$ of typical values from TNG50 over a large range of galaxy masses.

\subsection{Comparisons to Recent Work}

We now highlight results from previous work on \MgII{} absorption in observations and simulations in the context of our results. Observations of \MgII{} using sight lines near the major axis of galaxies have generally found that gas is corotating with the galaxy both for small impact parameters of $< 15 \; \rm{kpc}$ \citep[e.g.,][]{Bouche16} and large impact parameters of $> 50 \; \rm{kpc}$ \citep[e.g.,][]{MartinC19}. Using a lensed system, \cite{Lopez20} observed multiple sight lines of the same CGM and measured a decreasing \MgII{} rotation curve that is qualitatively similar to Figure \ref{f:vprofiles}. However, their absorption data only go out to $\approx30 \; \rm{kpc}$. Our work suggests \MgII{} rotation curves should continue to decrease to at least 100 kpc, though based on the maps in Figure \ref{f:MgII_maps} the \MgII{} column densities at those distances are significantly below current observational limits.

While this paper is focused on \MgII{} gas near the major axis, there are also recent results suggesting \MgII{} outflows along the minor axis of galaxies with velocities $> 100 \; \rm{km \; s^{-1}}$ \citep[e.g.,][]{Schroetter19,Zabl20}. It is worth noting though that \cite{Mortensen21} found a lensed system with \MgII{} on the geometric minor axis of the absorber galaxy with LOS velocities $< 100 \; \rm{km \; s^{-1}}$ and a large velocity dispersion, indicating that the kinematics of \MgII{} outflows may vary significantly. We showed in Figures \ref{f:covering} and \ref{f:vprofiles} that \MgII{} absorption along the minor axis is weaker than along the major axis, and that there are no net \MgII{} outflows along the minor axis in the TNG fiducial sample. This result appears to be discrepant with the previously cited observational papers, but we defer a detailed analysis to a future paper.

\cite{Ho20} recently studied similar aspects of \MgII{} absorption in the EAGLE simulation at $z\approx 0.3$ and found results broadly consistent with ours. Specifically, they measure a rotating \MgII{} structure around star-forming galaxies as well as a lower detection fraction of \MgII{} near the minor axis. They also find that higher-mass galaxies host detectable (i.e.,~above a fixed column density) \MgII{} structures out to larger distances in the CGM, which we indirectly show with the EW distributions in Figure \ref{f:mass_and_resolution}, where higher-mass halos have more strong absorbers.

\section{Summary} \label{sec:summary}

We have simulated \MgII{} absorption in the CGM of halos from TNG50 comparable to the major-axis sight lines observed in the MEGAFLOW survey by \cite{Zabl19} and compared absorption and kinematic properties of the two samples. We also examined the 3D kinematics of the \MgII{} in TNG50. Our conclusions are as follows:

\begin{enumerate}
    \item The equivalent widths of absorber-selected halos (i.e.,~strong absorbers) from TNG50 match reasonably well with the equivalent widths of major-axis sight lines from \cite{Zabl19} (Figure \ref{f:EW_vs_b}). 
    \item A majority of halos are strong absorbers at the smallest impact parameter studied (15 kpc), but the strong absorber fraction drops quickly as a function of distance (Figure \ref{f:EW_vs_b}).
    \item The stacked velocity spectra of TNG50 strong absorbers match the stacked spectra of \citet{Zabl19} very well, thus supporting the physical interpretation of corotation both below 30 kpc, where the spectra are strongly peaked near $\sim0.5 V_{\rm{vir}}$ and symmetric, and above 30 kpc, where the spectra are similarly peaked but are much noisier, broader, and asymmetric (Figure \ref{f:spectra}). 
    \item In TNG50, \MgII{} gas has velocity profiles nearly identical to gas below a temperature cutoff of $3\times10^4 \; \rm{K}$, meaning \MgII{} absorption is a good proxy for cold gas kinematics in general. There is substantial rotation and typical inflow velocities of up to $50 \; \rm{km} \; \rm{s^{-1}}$ out to $\sim40 \; \rm{kpc}$ in the CGM (Figure \ref{f:vprofiles}).
    \item The radial and polar velocity components by themselves do not cause any net velocity shift in the stacked spectrum, which implies that \MgII{} absorption kinematics alone cannot be used to measure typical inflow speeds of rotating gas in the CGM. (Figure \ref{f:vcomponents}). 
    \item \MgII{} absorption strengths and spectra are stronger and broader for halos more massive than the fiducial sample of $10^{11.5}-10^{12} \; M_{\odot}$ halos but do not change very much for halos less massive than the fiducial sample. Lowering the resolution from TNG50 to TNG100 only modestly changes any of the \MgII{} kinematic properties (Figure \ref{f:mass_and_resolution}).
    \item The median-specific angular momentum of the \MgII{} component of the CGM as a function of galactic stellar mass is very similar to that of both cold and hot CGM gas, and it is larger than that of the dark matter halo and the stars in the galaxy by $\sim0.2 \; \rm{dex}$ and $\sim0.8 \; \rm{dex}$, respectively. Estimates of the specific angular momentum of \MgII{} from the \citet{Zabl19} data are also reasonably close to the values from TNG50 to within a factor of $\sim0.5 \; \rm{dex}$. (Figure \ref{f:jM}).
\end{enumerate}

This work demonstrates that generating mock \MgII{} observations from TNG50 generates absorption spectra that are comparable to real data. In particular, our results are consistent with the emerging picture of rotating \MgII{} gas found in observations and also other simulations. In future work, we plan to widen our investigation to include other ions that trace warmer and more diffuse gas, as well as follow gas at particular redshifts backward and forward through time to determine the stability of various ion structures and their role in transporting angular momentum to or from the galaxy. 

\acknowledgments

We thank Johannes Zabl, \'Edouard Tollet, Joakim Rosdahl, and J\'er\'emy Blaizot for insightful and useful discussions, as well as Cameron Hummels for assistance with \textsc{Trident}. We also thank the anonymous referee for helpful comments. D.D. acknowledges support from the Chateaubriand Fellowship of the Office for Science \& Technology of the Embassy of France in the United States. N.B. acknowledges funding support from the French Agence National de la Recherche (ANR) grant ``3DGasFlows'' (ANR-17-CE31-0017). Flatiron Institute is supported by the Simons Foundation. G.L.B. acknowledges financial support from the NSF (grants AST-1615955, OAC-1835509) and computing support from NSF XSEDE. F.M. acknowledges support through the Program ``Rita Levi Montalcini'' of the Italian MUR.

\software{\textsc{Trident} \citep{Hummels17},  
          \textsc{yt} \citep{Turk11},
          \textsc{Cloudy} \citep{Ferland13}, 
          \textsc{NumPy} \citep{vanderWalt11}, 
          \textsc{Matplotlib} \citep{Hunter07}, 
          and \textsc{IPython} \citep{Perez07}
          }

\bibliography{references}{}
\bibliographystyle{aasjournal}

\end{document}